\documentclass[times,5p,twocolumn,final]{elsarticle}
\newtheorem{definition}{Definition}
\usepackage{amssymb}
\usepackage[title]{appendix}
\usepackage{amsmath}
\usepackage[inline]{enumitem}
\usepackage{caption}
\usepackage{subcaption}
\usepackage{mathtools}
\usepackage{footmisc}
\usepackage{hyperref}
\usepackage{booktabs}
\usepackage{xcolor}
\usepackage{balance}
\usepackage[boxruled,linesnumbered]{algorithm2e}
\definecolor{boxgrey}{HTML}{F3F3F3}

\newcounter{subdefinition}[definition]
\renewcommand{\thesubdefinition}{\thedefinition.\arabic{subdefinition}}

\def\threedigits#1{\number#1}
\newcommand{\hlbox}[2]{
  \begin{center}
    \fcolorbox{white}{boxgrey}{
      \parbox{.9\columnwidth}{\noindent \textbf{#1}. \textit{#2}}
    }
  \end{center}
}
\journal{Knowledge-Based Systems}

\begin{document}

\begin{frontmatter}

\title{Reinforcement Recommendation Reasoning through Knowledge Graphs for \\ Explanation Path Quality}

\author{Giacomo Balloccu}
\author{Ludovico Boratto}
\author{Gianni Fenu}
\author{Mirko Marras}

\affiliation{organization={Department of Mathematics and Computer Science, University of Cagliari},
            addressline={\\Via Ospedale 72}, 
            city={Cagliari},
            postcode={09124}, 
            country={Italy}}

\begin{abstract}
Numerous Knowledge Graphs (KGs) are being created to make Recommender Systems (RSs) not only intelligent but also knowledgeable. 
{\color{black} Integrating a KG in the recommendation process allows the underlying model to extract reasoning paths between recommended products and already experienced products from the KG. These paths can be leveraged to generate textual explanations to be provided to the user for a given recommendation. However, the existing explainable recommendation approaches based on KG merely optimize the selected reasoning paths for product relevance, without considering any user-level property of the paths for explanation.}
In this paper, we propose a series of quantitative properties that monitor the quality of the reasoning paths {\color{black} from an explanation perspective}, based on recency, popularity, and diversity. 
We then combine in- and post-processing approaches to optimize for both recommendation quality and reasoning path quality.
Experiments on three public data sets show that our approaches significantly increase reasoning path quality according to the proposed properties, while preserving recommendation quality. Source code, data sets, and KGs are available at \url{https://tinyurl.com/bdbfzr4n}. 
\end{abstract}


\end{frontmatter}

\section{Introduction}

\vspace{1mm} \noindent{\bf Motivation.} 
Explaining to users {\em why} certain results have been provided to them has become an essential property of modern systems. 
Regulations, such as the European General Data Protection Regulation (GDPR), call for a ``right to explanation'', meaning that, under certain conditions, it is mandatory by law to generate awareness for the users on how a model behaves~\cite{GoodmanF17}. 
Explanations have been also proved to have benefits from a business perspective, by increasing trust in the system, helping the users make a decision faster, and persuading a user to try and buy~\cite{Tintarev2007}.
Recommender Systems (RSs) are a notable class of decision-support systems that urge supporting explanations.
Existing RSs often act as black boxes, not offering the user any justification for the provided recommendations. 
Efforts have been devoted to challenge these black boxes to make recommendation a transparent social process~\cite{XianFMMZ19}.

\vspace{1mm} \noindent{\bf State of the Art.} {\color{black} 
Transparency has been increasingly recognized as an essential yet prominent objective by the machine-learning research community.
This importance has led to a proliferation of interpretability and explainability methods. 
Notable methods rely for instance on the use of bi- and tri-partite graphs \cite{yang2021interpretable,10.1016/j.knosys.2021.107922,10.1016/j.knosys.2022.108274} and Knowledge Graphs (KGs) \cite{cao-etal-2018-neural, 10.1145/2926718}.
Integrating this external knowledge has resulted in both increased transparency and higher effectiveness in many domains.
First steps towards improving transparency in RS have been made by augmenting traditional recommendation models, that originally modelled user-product interactions only, with external knowledge about the users and the products. 
For its integration, prior work has adopted regularized and path-based approaches.} 

{\color{black} Regularized approaches~\cite{CaoWHHC19, CKE10.1145/2939672.2939673} extend the original objective function with a term that 
implicitly encodes high-order relations between users and products in the KG. 
Given a range of pre-defined user and product characteristics, the approaches belonging to this class compute a weight for each characteristic based on its importance for the recommendation of that product. 
To compute these weights, some methods aggregate the neighbours of a certain user or product, as an example \cite{Wang00LC19, ripple-net/10.1145/3269206.3271739}.
Other methods require ad-hoc model-dependent modules for importance weights computation \cite{yang2021interpretable}.
These importance weights can be used to select a pertinent textual explanation from a fixed set of explanations pre-defined by scientists.
For instance, \cite{10.1016/j.knosys.2021.107922} defined a user-centered set (e.g., "users living in the same city") and a product-centered set (e.g., "news with the same topic") of possible explanations. In case the users' city was found to have the highest weight, the textual explanation "users living in the same city" was shown to the user. 
However, these approaches suffer from several limitations, e.g., high model dependence, hand-crafted and narrowed set of rules, and lack of specificity.}

\begin{figure*}
\centering
\includegraphics[width=1.0\textwidth]{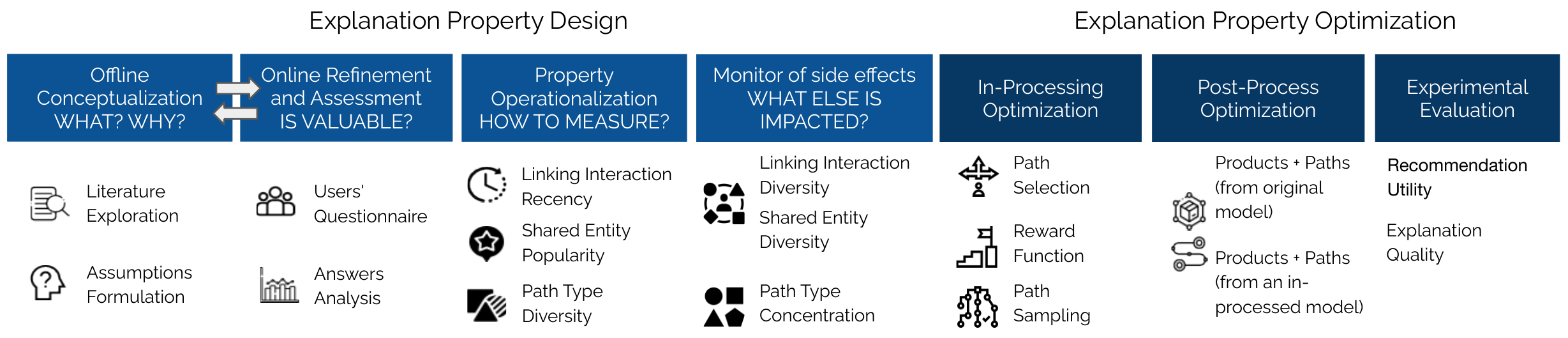}
\caption{We adopted a mixed approach to analyze the space of relevant reasoning path quality properties comprehensively. As a result, we identified and operationalized three reasoning path quality metrics. Explainable paths and the resulting lists of recommended products were optimized for these metrics through in- and post-processing approaches. We finally assessed recommendation quality and reasoning path quality.} 
\label{fig:path-img}
\vspace{4mm}
\end{figure*}

Path-based approaches instead rely on pre-computed paths (tuples) that model the high-order relations between users and products, according to the KG structure~\cite{Wang_Wang_Xu_He_Cao_Chua_2019, 10.1145/3219819.3219965,10.1109/TKDE.2018.2833443}.
These tuples serve as an additional input to the recommendation model during training. 
Compared to regularized approaches, the approaches in this second class identify candidate reasoning paths between the recommended product and already experienced products. 
These paths can be then leveraged to instantiate an explanation template or as an input to text generation techniques, to finally obtain a textual explanation. 
In the movie domain, a path between a movie already watched by the user ($movie_1$) and a movie recommended to that user ($movie_2$), shaped in the form of ``$user_1$ \texttt{watched} $movie_1$ \texttt{directed} $director_1$ \texttt{directed$^{-1}$} $movie_2$'', can be used to provide the template-based textual explanation ``$movie_2$ is recommended to you because you watched $movie_1$ also $directed$ by $director_1$''.
Path-based approaches address several limitations observed for regularized approaches.

\vspace{1mm} \noindent{\bf Open Issues.} 
Due to the large amount of nodes and edges in the KG and the inordinate computational resources required to explore it fully, path-based approaches generally perform the path extraction and the recommendation steps separately. 
Considering that the path extraction is not directly optimized for the recommendation objective, the selected reasoning paths do not often encode the true functioning of the model and serve only to generate sub-optimal post-hoc explanations.  
To understand the underlying model reasoning, the need of path-based approaches that can yield both the recommendation and its explanation (self-explainable) from a model jointly optimized for recommendation quality and reasoning path quality is timely.

State-of-the-art path-based approaches \cite{XianFMMZ19, song2022ekar, 10.1145/3340531.3412038, 10.1145/3404835.3462847, 10.1145/3485447.3512083,10.1145/3447548.3467315} rely on a Reinforcement Learning (RL) agent, conditioned on the user and trained for navigating to potentially relevant products for that user through path reasoning.
The agent then samples paths between the user and the products in the KG, by simultaneously conducting product recommendation and
path selection, based on the probability the agent took an existing path to reach a given recommended product. 
{\color{black} However, to select the user-product paths for textual explanation, all the existing path-based approaches merely consider an inner-functioning probability.
None of them embeds objectives pertaining to how the selected path and the derived textual explanation are perceived by users (e.g., a user might prefer explanations linked to more recently experienced products). From a technical perspective, this limitation means that no existing path-based method includes optimization terms for user-level properties of the paths from an explanation perspective, but just a term that measures the extent to which a product is relevant for a user.
Hence, approaches that can optimize the selected paths for explanation-related properties from the user perspective are urging.} 

\vspace{1mm} \noindent{\bf Our Approach and Contributions.} 
{\color{black} In this paper, we aim to achieve the above mentioned objective on template-based textual explanations, focusing on user-level offline properties of the reasoning paths used for instantiating templates.
Given that such paths are the input of the textual explanation generation, optimizing the selected paths for properties relevant for the user can improve the perceived quality of the textual explanations. }

However, there is a potentially large set of path-related properties to consider. 
Our study in this paper explores the space of relevant reasoning path properties through a mixed-method approach.
We combine both literature review (also including psychological dimensions) and user's questionnaires (investigating which and whether users perceive certain properties as valuable). 
From this analysis, we identified a set of properties deemed as important by users, namely recency, popularity {\color{black}(connected with novelty and serendipity)~\cite{10.1145/3459637.3482394}}, and diversity~\cite{DBLP:journals/tiis/KaminskasB17}.
Considering a single explanation, recent linking interactions might lead to a lower memory overload for users (to link back to that past interaction) {\color{black}and textual explanations better connected with their recent tastes.}
Popular entities might be already known by the user and, possibly, lead to a better understanding of the provided explanation. Conversely, niche entities might help users learn novel links across products in that domain. 
Finally, given that recommendations are often provided as a list, with a textual explanation for each recommended product, the explanation quality also depends on how explanations are perceived as a whole over the list.  
{\color{black}Due to the imbalances in the KG, existing approaches tend to mostly produce collaborative filtering reasoning paths that lead to a textual explanation in the form ``...because another user that watched $movie_1$ liked $movie_2$''. 
This type of explanation may be perceived as naive and vague, and hence negatively impact the trust towards the system.  
Considering the \emph{diversity} of path type (e.g., directed or starring) can lead to better perceived textual explanations.}

{\color{black} Integrating these user-level optimization objectives can support the generation of better perceived explanations for recommendation in several applicative areas.
For instance, in the music domain, providing explanations related to the last listened songs (recency) might lead to the user feeling that the system is better understanding their interests in the current session. 
In the book domain, including popular entities in the generated textual explanation (popularity) has the potential to increase their relevance and pertinence. 
As another example, in online course recommendation, providing a more diverse set of explanation types might increase user trust on what the system found relevant for recommending that course. 
Similar observations can be made on other recommendation domains (e.g., movies, products, and points of interest), highlighting that our contribution is general and flexible enough for a broad range of scenarios.   
}

The recognized importance for these properties motivated us to operationalize a range of metrics for recency, popularity, and diversity of explanation. 
We proposed in- and post-processing approaches that optimize the recommended products and the accompanying textual explanations for these metrics.
We assessed the impact of our approaches on recommendation quality and explanation path quality, investigating whether any trade-off aroused. 
Figure \ref{fig:path-img} summarizes our pipeline. 
Concretely:

\begin{enumerate}

\item We introduce reasoning path quality metrics to measure recency, popularity, and diversity of explanation, and explore the extent to which the paths generated by existing approaches hold these properties.

\item We propose and combine a suite of in- and post-processing approaches, acting on both the recommended products and the reasoning paths, optimizing for the proposed reasoning path quality metrics. 

\item We show the benefits and any side-effect of our approaches on the trade-off between recommendation quality and reasoning path quality, on three real-world public data sets, against seven baselines.
\end{enumerate}

\section{Problem Formulation}\label{sec:preliminaries}
We introduce the notation adopted in our study (summarized in Table~\ref{tab:table-of-notation}) and then define the addressed task.

\begin{table}[!b]
  \vspace{1mm}
  \caption{Mathematical notation adopted in our study.}
  \label{tab:table-of-notation}
  \centering
\resizebox{1\linewidth}{!}{
  \begin{tabular}{ll}
    \toprule
    \textbf{Symbol} & \textbf{Description} \\
    \hline
    \midrule
    $\mathcal G$ & Knowledge graph \\
    $\mathcal E$ & Entities set in a knowledge graph\\
    $\mathcal R$ & Relations set in a knowledge graph \\
    $\mathcal U$ & User entities set ($\mathcal U \subset E$) \\
    $\mathcal P$ & Product entities set ($\mathcal P \subset E$) \\
    $\mathcal P_u$ & Product entities set the user $u$ previously interacted with ($\mathcal P_u \subset P$) \\
    $\mathcal F$ & User-product feedback matrix\\
    \hline
    $k$ & Reasoning path length ($k \in \mathbb N$) \\
    $\mathcal L^k$ & Reasoning paths set of length $k$ \\
    $\mathcal L^k_{u}$ & Candidate reasoning paths of length $k$ between user $u \in \mathcal U$ and $\forall$ product $p \in \mathcal P$\\
    $\mathcal L^k_{u,p}$ & Candidate reasoning paths set of length $k$ between user $u \in \mathcal U$ and product $p \in \mathcal P$ \\
    \hline
    $\theta$ & Recommendation model parameters set\\
    $\Tilde{\theta}$ & Parameters set for an optimized recommendation model\\
    n & Recommended list size ($n \in \mathbb N$)\\
    $\Tilde{\mathcal P}^n_u$ & Recommended products list of size $n$ for user $u \in \mathcal U$\\
    $\Tilde{\mathcal L}^k_{u}$ & Set of reasoning paths of length $k$ selected to explain the recommended products $\Tilde{\mathcal P}^n_u$ of user $u \in \mathcal U$\\
    $\Tilde{\mathcal F}_{u,p}$ & The predicted relevance of product $p \in \mathcal P$ for user $u \in \mathcal U$\\
    $\Tilde{\mathcal S}_{l^k_{u,p}}$ & The predicted probability the path $l^k_{u,p}$ of length $k$ is followed from user $u \in \mathcal U$ to product $p \in \mathcal P$\\
    \hline
    $\Upsilon$ & Template-based textual generation function \\
    $\Tilde{\Upsilon}$ & Specific template-based textual generation function used in our study\\
    $\mathcal V$ & Vocabulary of words admitted for generating a template-based textual explanation\\
    \bottomrule
  \end{tabular}}
\end{table}

\subsection{Knowledge Graph Definition}

We first formalize the concept of knowledge graph we used.

\begin{definition} {{\color{black} (Knowledge Graph)} A knowledge graph is defined as a set of triplets $\mathcal G = \{(e_h, r, e_t) | \; e_h, e_t \in \mathcal E, r \in \mathcal R\}$ where $\mathcal E$ is the set of entities and $\mathcal R$ is the set of relations connecting two entities. Each triplet $(e_h, r, e_t) \in \mathcal G$ models a relation $r \in \mathcal R$ between a head entity $e_h \in \mathcal E$ and a tail entity $e_t \in \mathcal E$.}
\end{definition}

{\color{black} Following prior work in explainable recommendation \cite{ripple-net/10.1145/3269206.3271739, XianFMMZ19}, we consider a knowledge graph where at least two types of entities are present: the set $\mathcal U \subset \mathcal E$ of users and the set $\mathcal P \subset \mathcal E$ of products. 
The special relation $r_f \in \mathcal R$ between a user and a product models the user feedback and is dependent on the domain (e.g., a user ``watched'' a movie or ``listened to'' a song).} 
Example additional entities (and relations) might be actors (an actor ``starred'' a movie) or directors (a director ``directed'' a movie) in the movie domain or artists (an artist ``interpreted'' a song) and producers (a producer ``produced'' a song) in the music domain. 
{\color{black} Based on this formalization, a k-hop reasoning path between two entities in $\mathcal G$ can be formalized as follows.}

\begin{definition} {{\color{black} (K-Hop Reasoning Path)} A k-hop path between entities $e_0, e_k \in \mathcal E$ is defined as a sequence of $k+1$ entities connected by $k$ relations and denoted as $l^k_{e_0, e_k} = \{ e_0 \xleftrightarrow[]{r_1} e_1 \xleftrightarrow[]{r_2} ... \xleftrightarrow[]{r_k} e_k$\}, where $e_{i-1} \xleftrightarrow[]{r_i} e_i$, with $0 < i \leq k$, is assumed to represent $(e_{i-1}, r_i, e_i)$ $\in \mathcal G$ or its inverse $(e_i, r_{i}^{-1}, e_{i-1})$ $\in \mathcal G$.}
\end{definition}

{\color{black} We refer to the relation $r \in \mathcal R$ either as \emph{forward} (active) if $(e, r, e')$ $\in \mathcal G$ and $e \rightarrow e'$ or backward (passive) if $(e, r^{-1}, e')$ $\in \mathcal G$ and $e \leftarrow e'$.
We denote as $\mathcal L^k_{u}$ the set of all possible $k$-hop paths between a user $u$ and any product.

In the movie domain, an example $3$-hop path between a user and a movie might be ``$user_1$ \texttt{watched} $movie_1$ \texttt{directed}$^{-1}$ $director_1$ \texttt{directed} $movie_2$''.
Each relation $r \in \mathcal R$ uniquely identifies the candidate sets to be used for the head and tail entities (e.g., the actor and movie sets for the relation ``starred'' or the artist and song sets for the relation ``interpreted'').

We finally introduce the type of a reasoning path as follows:

\begin{definition} {{\color{black} (Reasoning Path Type)} Given a k-hop reasoning path $l^k_{e_0, e_k} = \{ e_0 \xleftrightarrow[]{r_1} e_1 \xleftrightarrow[]{r_2} ... \xleftrightarrow[]{r_k} e_k$\}, the type of the path $l^k_{e_0, e_k}$ is the last relation $r_k \in \mathcal R$.}
\label{def:path-type}
\end{definition}

The type of the example path is therefore ``directed''.

\subsection{Recommendation over Knowledge Graphs} \label{sec:rec-kg}
We define the user-product feedback $\mathcal F \in \mathbb R^{|\mathcal U| * |\mathcal P|}$ as a function, with $\mathcal F(u, p) = 1$ in case user $u$ interacted with product $p$, $\mathcal F(u, p) = 0$ otherwise. 

Given this matrix, a traditional recommendation model not explicitly using KGs aims to estimate relevances $\Tilde{\mathcal F}(u, p) \in [0, 1]$ of unobserved entries in $\mathcal F$ and use them for ranking products. 
This operation can be abstracted as learning a model $\theta : (\mathcal U, \mathcal P) \rightarrow \mathbb R$. 
Products are sorted by decreasing relevance for a user, and the top-$n$ ($n \in \mathbb N$), products $\Tilde{\mathcal P}^n_u$ are recommended.  

Being interested in improving recommendation transparency, given a certain $k \in \mathbb N$, our focus is to recommend to user $u$ a useful set of products $\Tilde{\mathcal P}^n_u$, where every product $p \in \Tilde{\mathcal P}^n_u$ is associated with a reasoning path $\Tilde{l}^k_{u,p} \in \Tilde{\mathcal L}^k_{u}$ to be used as an input for the textual explanation generation. This path is selected by the model as the most representative for the recommended product $p$ among the set of all the predicted $k$-hop paths $\mathcal L^k_{u,p}$ between user $u$ and product $p$. 
Our addressed task, named as Knowledge Graph Reasoning for Explainable Recommendation (KGRE-Rec), can be hence formalized as:}

\begin{definition}
\label{def:kbre-problem}
{{\color{black} (Knowledge Graph Reasoning for Explainable Recommendation Task)} Given a knowledge graph $\mathcal G$ and an integer $K \in \mathbb N$ representing the maximum hop length, the goal is learning a model $\theta : (\mathcal U, \mathcal P, \mathcal L^k) \rightarrow \mathbb R^2$ able to estimate (i) the point-wise relevances $\Tilde{\mathcal F}(u, p) \in [0, 1]$ for unobserved entries in $\mathcal F$ and (ii) the probabilities $\Tilde{\mathcal S}(l_{u,p}) \in [0, 1]$ that a path $l_{u,p} \in \mathcal L^k_{u,p}$ is followed to reach product $p$ from user $u$ in $\mathcal G$, with $2 < k \leq K$.}
\end{definition}

Given a user $u$, the products are sorted by decreasing relevance for user $u$ based on $\Tilde{\mathcal F}$, and the top-$n$ products $\Tilde{P}^n_u$ are recommended. 
For each product $p \in \Tilde{P}^n_u$, the predicted paths $\Tilde{\mathcal L}^k_{u,p}$, with $2 < k \leq K$, from user $u$ to product $p$ are sorted by decreasing probability of being followed based on $\Tilde{\mathcal S}$, and the top path $\Tilde{l}^k_{u,p}$ is selected to generate the textual explanation.

\subsection{Recommendation Objectives Definition}
We believe that selecting the reasoning path based on an inner-functioning probability does not necessarily lead to high-quality explanations from the user's perspective.
We therefore assume that there exist (i) a set $\mathcal C^{RQ}$ of recommendation quality properties, denoted with functions in the form $ \mathcal P^* \xrightarrow{} \mathbb R$, and (ii) a set $\mathcal C^{EQ}$ of user-related reasoning path properties, denoted with functions in the form $\mathcal L^* \xrightarrow{} \mathbb R$.
An example reasoning path property might be the \emph{recency} of the prior interaction attached to the reasoning path, i.e., {\color{black} how recent the ``watched'' interaction between $user_1$ and $movie_1$ in the example path ``$user_1$ \texttt{watched} $movie_1$ \texttt{directed}$^{-1}$ $director_1$ \texttt{directed} $movie_2$'' is.}
An ideal explainable RS $\tilde{\theta}$ would consider both perspectives, maximizing the following objective function:

\vspace{-2mm}
\begin{equation}
\tilde{\theta} = \underset{\theta}{\operatorname{argmax}} \mathop{\mathbb{E}}_{u \; \in \; \mathcal U} \; \sum_{c \; \in \; \mathcal C^{RQ}} c(\Tilde{P}^n_u) + \sum_{c \; \in \; \mathcal C^{EQ}} c(\Tilde{\mathcal L}^k_u)
\label{eq:problem-definition}
\end{equation}
\vspace{1mm}

For simplicity, we assume that recommendation and reasoning path quality are equally weighted, leaving user's specific weights as a future work.
{\color{black} Given the heterogeneous nature of the reasoning path properties, prior work assumed that the original recommendation model is optimized only on recommendation quality ($C^{RQ}$) and that reasoning path quality ($\mathcal C^{EQ}$) is optimized via post-processing \cite{10.1145/3477495.3532041}. 
Our study in this paper explores the in-processing optimization of both recommendation quality and reasoning path quality, and the combination of in- and post-processing optimization. } 

\subsection{Textual Explanation Generation}
{\color{black}
The reasoning paths selected for the recommended products serve as an input to the textual explanation generation step. 
Such generation step may be based, for instance, on templates or advanced natural language generation. 
In our study in this paper, we focus on template-based textual explanation generation according to the reasoning paths selected by an underlying model \cite{10.1145/2600428.2609579}. 
Generally, a template can be seen as a string literal that includes expressions whose content is produced by a model. 
Formally:}

\begin{definition}
\label{def:explanation-template}
{{\color{black} (Explanation Template)} Given a vocabulary of words $\mathcal V$, an explanation template is abstracted as a function $\Upsilon : \mathcal L \rightarrow \mathcal V^*$ that, given a path, produces a string representing a human-readable textual explanation.}
\end{definition}

For each recommended product $p \in \Tilde{P}^n_u$, the path $\Tilde{l}^k_{u,p}$ is selected to generate the textual explanation as described in Section \ref{sec:rec-kg}. 
Such path includes three conceptual parts: 
\begin{itemize}
\item \textit{Linking Interaction} $(e_0 = u \in \mathcal U, r_1 = r_f, e_1 = p_1 \in \mathcal P)$ of a user $u$ with a product $p_1$ according to the feedback information provided in $\mathcal F$.
\item \textit{Entity chain} $(e_{j-1}, r_{j-1}, e_j)$, with $j=2, \dots, k-1$, from the product $p_1 \in \mathcal P$, with $e_1 = p_1$, and connecting to non-product entities $e \notin \mathcal P$ (shared entities).
\item \textit{Recommendation} $(e_{k}, r_{k}, e_{k} = p_2 \in \mathcal P)$ that connects product $p_2$ to be recommended to the path to user $u$.
\end{itemize}

For instance, given the guiding example path, ($user_1$ \texttt{watched} $movie_1$) is the past interaction, ($movie_1$ \texttt{directed$^{-1}$} $director_1$) is the entity chain, and ($director_1$ \texttt{directed} $movie_2$) is the recommendation. 
Based on this conceptualization, we formally define the specific concept of $k$-hop explanation template adopted throughout our study, as follows:

\begin{definition}
\label{def:reasoning-path-template}
{{\color{black} (K-Hop Explanation Template)} Given a vocabulary of words $\mathcal V$, a $k$-hop explanation template is abstracted as a function $\Tilde{\Upsilon}: \mathcal L^k \rightarrow \mathcal V^*$ and implemented as $\Tilde{\Upsilon}(l^k_{e_0, e_k} = \{ e_0 \xleftrightarrow[]{r_1} e_1 \xleftrightarrow[]{r_2} ... \xleftrightarrow[]{r_k} e_k\})$ = "$<e_k>$ is recommend to you because you $<r_1>$ $<e_1>$ also $<r_k>$ by $<e_{k-1}>$”.}
\end{definition}

{\color{black}Considering the example path, through our definition of $\Tilde{\Upsilon}$, we can produce the textual explanation ``$movie_2$ is recommended to you because you \texttt{watched} $movie_1$ also \texttt{directed} by $director_1$''.}

\section{Explanation Property Design}\label{sec:exp-properties}
{\color{black} Defining relevant reasoning path properties and optimizing a recommendation model for them are the two key tasks emerged from our formalization.
In this section, we delve into the first task, by describing the set of reasoning path properties investigated later on in our study. 
The identified properties consider three key aspects pertaining to reasoning paths, connected with the recency of the linking interaction, the popularity of the shared entity, and the diversity of explanation path types.
{\color{black}The relevance and importance of these aspects for users emerged from our mixed-method study, which capitalized on both literature analysis and user questionnaires\footnote{{\color{black}A questionnaire copy is available at \url{https://tinyurl.com/exp-quality-survey}}.} under a paired-preference protocol \cite{10.1145/3477495.3532041}.}
For each aspect, we present here the motivations emerged from both the literature analysis and the users assessment. We then provide a mathematical formulation and practical examples for each derived metric, using the toy path ``$u$ \texttt{listened} $song_1$ \texttt{featured}$^{-1}$ $artist_1$ \texttt{featured} $song_2$''.}

\subsection{Recency of the Linking Interaction}
The first emerged aspect was the recency of an interaction with the already experienced product in a path, i.e., $u$ \texttt{listened} $song_1$ in the toy path. 

\vspace{1mm}\noindent\textbf{Motivation.} 
From the literature analysis, it emerged that incorporating the time of interaction into recommendation models is a practice that has led to gains in recommendation quality~\cite{LeePP08,DingLO06}, and could influence how users will perceive the final explanation as well.
Indeed, an explanation related to a recent interaction would be intrinsically easier to catch for a user, while older interactions might not be perceived as valuable or might even be not remembered by the users.
As an example, we might consider a user of a movie platform who was highly active in the past, but inactive in the last years. 
This user starts to use again the platform, they perform various interactions with new movies, and start receiving new recommendations. 
So, rewarding an explanation based on the freshness of the interaction would be useful. 
Fresher interactions could also be easier to understand and would be more timely, compared to those with products associated to very old interactions. 
The importance of such property was also confirmed in the outcomes of our users assessment. 
Indeed, we observed that $64.6\%$ of the participants preferred to see an explanation involving a product closely experienced in time, $6.8\%$ opted for explanations involving older interactions, and the remaining $28.6\%$ of the participants said that this property would not be relevant for them.

\vspace{1mm}\noindent\textbf{Operationalization.} 
To operationalize this property, we considered the time since the linking interaction in the reasoning path occurred. 
Given a user $u \in \mathcal U$ and the set $\mathcal P_u$ of products user $u$ interacted with, we denote the list of $u$'s interactions, sorted chronologically, by $\texttt{T}_u = [(p^i, t^i)]$, where $p^i \in \mathcal P_u$ is a product experienced by user $u$, $t^i \in \mathbb N$ is the timestamp that interaction of user $u$ with product $p_i$ occurred, and $t^i \leq t^{i+1}$ $\forall i = 1, \dots, |\mathcal P_u|$. 
We applied an exponentially weighed moving average to the timestamps in $\texttt{T}_u$, to obtain the recency score of each interaction performed by user $u$. Formally:

\begin{definition}
{{\color{black}(Interaction Recency - IR)} Given a user $u \in \mathcal U$, a chronologically sorted list of $u$'s interactions $\texttt{T}_u$, and a time decay $\beta_{IR} \in [0, 1]$, the recency of the interaction $(p^i, t^i) \in \texttt{T}_u$ is defined as:}
\vspace{-2mm}
\begin{equation} 
IR(p^i, t^i) = ( 1 - \beta_{IR} ) \; IR(p^{i-1}, t^{i-1}) + \beta_{IR} \; t^i \; \text{\bf{with}} \; IR(p^1, t^1) = t^1     
\end{equation}
\end{definition}

The $IR$ values were min-max normalized among those of all the interactions in $\texttt{T}_u$ to lay in the range $[0, 1]$.
Given this formalization, we can introduce the recency of the linking interactions over a list of reasoning paths.   

\begin{definition}
{{\color{black}(Linking Interaction Recency - LIR)} Given a user $u$, a chronologically sorted list of $u$'s interactions $\texttt{T}_u$, and the set of reasoning paths $\Tilde{\mathcal L}^k_u$ selected for explaining the recommended products $\tilde{\mathcal P}_u$, with $|\Tilde{\mathcal L}^k_u|=|\tilde{\mathcal P}_u|$, the linking interaction recency over the selected reasoning paths is defined as:}
\vspace{-2mm}
\begin{equation}
LIR(\Tilde{\mathcal L}^k_u) = \frac{1}{|\Tilde{\mathcal L}^k_u|} \sum_{l = \{ e_0 \xleftrightarrow[]{r_1} e_1 \xleftrightarrow[]{r_2} ... \} \; \in \; \Tilde{\mathcal L}^k_u}{IR(e_1, t)}  \; \text{\bf{with}} \; (e_1, t) \in \texttt{T}_u \;  
\label{eq:lir}
\end{equation}
\end{definition}

In case the LIR values are close to $0$ ($1$), the linking interaction of the selected reasoning path is on average old (recent). 

{\color{black} Optimizing the set of reasoning paths $\Tilde{\mathcal L}^k_u$ selected for explaining the recommended products for LIR can however introduce possible side effects. 
Indeed, it might be possible that all or the majority of the selected reasoning paths focus on a tiny set of recent linking interactions.
The very extreme case would be that all the selected reasoning paths include the same most recent linking interaction.
This effect could make the link to already experienced products for the provided explanations repetitive and impact on the perceived explanation quality, as an example. 
Due to this reason, we decided to operationalize an additional related metric for monitoring the diversity of the linking interactions included in the selected reasoning paths (i.e., how many different interactions are linked to provided explanations)}. Formally:

\begin{definition}
{{\color{black}(Linking Interaction Diversity - LID)} Given a user $u$ and the set of reasoning paths $\Tilde{\mathcal L}^k_u$ selected for explaining the recommended products $\tilde{\mathcal P}_u$, with $|\Tilde{\mathcal L}^k_u|=|\tilde{\mathcal P}_u|$, the linking interaction diversity over those paths is:}
\vspace{-2mm}
\begin{equation}
LID(\Tilde{\mathcal L}^k_u) = \frac{|\{e_1 \; | \; \forall l = \{ e_0 \xleftrightarrow[]{r_1} e_1 \xleftrightarrow[]{r_2} ... \} \; \in \; \Tilde{\mathcal L}^k_u \}|}{|\Tilde{\mathcal L}^k_u|}
\label{eq:lid}
\end{equation}
\end{definition}

$LID$ values lay in the range $(0, 1]$, with values close to $0$ ($1$) meaning that the recommended list has a low (high) linking interaction diversity. 

\subsection{Popularity of the Shared Entity}
The second aspect identified in our study was related to the popularity of the shared entity, i.e., $artist_1$ in the example path. 

\vspace{1mm}\noindent\textbf{Motivation.} From our literature analysis, it emerged that, in traditional RS research, popularity is a concept generally connected with novelty (e.g., the less popular the recommended product among other users is, the higher the novelty is) and familiarity (e.g., the more the product is popular among other users, the higher the chance it will be familiar for the user). These two beyond-accuracy properties have been often recognized as important for the recommended products \cite{DBLP:journals/tiis/KaminskasB17}, according to the application scenario. {\color{black} We therefore considered to investigate the extent to which the popularity of the shared entity can influence the perceived quality of the explanation as well.} 
For instance, an artist who featured 20 songs might be considered more popular that one who featured 2 songs. 
In case a very unpopular recommended product is given, an explanation that contains a popular entity can help the user decide whether that product can be interesting for them. 
{\color{black} Moreover, it should be considered that the shared entity mentioned in an explanation can act as a source of context, since it can influence the perception of the usefulness of an product \cite{10.1145/2533670.2533675}}.
In \cite{SwearingenS02}, 70\% of the products that users expressed an interest in buying were familiar products.
These observations are also remarked in~\cite{PuCH12}, which considered the familiarity of the users with the recommended products. 
Conversely, in case the shared entity has a low popularity, the user may not catch the explanation, since they might not know that artist or actor presented in the explanation. 
Providing explanations associated with products that are too popular or redundant could however increase the filter bubbles in the explanations in the long term~\cite{GedikliJG14}.
Therefore, the popularity of the shared entity might be potentially minimized or maximized according to the overall strategy of the platform, e.g., promoting familiarity or novelty / serendipity. 
As a confirmation of this literature analysis, the answers to the second question of our users assessment showed that $63.3\%$ of users expressed an interest toward this property, i.e., $40\%$ ($24.3\%$) of the participants preferred a popular (unpopular) shared entity.
$35.7\%$ of the participants marked this property as not relevant.

\vspace{1mm}\noindent\textbf{Operationalization.} 
Motivated by our analysis, we operationalized a metric able to quantify the extent to which the shared entity included in a reasoning path is popular. 
We assume that the number of relations a shared entity is involved in the KG is a proxy of its popularity. 
For instance, the popularity of an actor is computed by counting how many movies that actor starred in.
We denote the list of entities participation of a given type $\lambda$ in the KG, sorted based on their popularity, by $\texttt{E}_\lambda = [(e^i, v^i)]$, where $e^i \in \mathcal E_\lambda$ is an entity of type $\lambda$, $v^i \in \mathbb N$ is the number of relations the shared entity $e_i$ is involved in (in-degree), and $v^i \leq v^{i+1}$ $\forall i = 1, \dots, |\texttt{E}_\lambda|$. 
We again apply an exponential decay to the number of relations in $\texttt{E}_\lambda$, to obtain the popularity score of an entity participation of type $\lambda$. Formally: 

\begin{definition}
{{\color{black}(Entity Popularity - EP)} Given an entity type $\lambda$, a popularity-sorted list of entities participation $\texttt{E}_\lambda$, and a popularity decay $\beta_{EP} \in [0, 1]$, the popularity of an entity participation $(e^i, v^i) \in \texttt{E}_\lambda$ is defined as:}
\vspace{-2mm}
\begin{equation} 
EP(e^i, v^i) = ( 1 - \beta_{EP} ) \; EP(e^{i-1}, v^{i-1}) + \beta_{EP} \; v^i \; \text{\bf{with}} \; EP(e^1, v^1) = v^1     
\end{equation}
\end{definition}

The $EP$ values were min-max normalized among those of all entities of a given type to lay in the range $[0, 1]$. 
Given this formalization, we can introduce the popularity of the shared entity over a list of reasoning paths.   

\begin{definition}
{{\color{black}(Shared Entity Popularity - SEP)} Given a user $u$, all popularity-sorted lists of entities participation $\texttt{E}_\lambda \; \forall \lambda$, and the set of reasoning paths $\Tilde{\mathcal L}^k_u$ selected for explaining the recommended products $\tilde{\mathcal P}_u$, with $|\Tilde{\mathcal L}^k_u|=|\tilde{\mathcal P}_u|$, the shared entity popularity over the selected reasoning paths is defined as:}
\vspace{-2mm}
\begin{equation}
SEP(\Tilde{\mathcal L}^k_u) = \frac{1}{|\Tilde{\mathcal L}^k_u|} \sum_{l = \{ e_0 \xleftrightarrow[]{r_1} ... \xleftrightarrow[]{r_k} e_k\} \; \in \; \Tilde{\mathcal L}^k_u}{EP(e_{k-1}, v)}  \; \text{\bf{with}} \; (e_{k-1}, v) \in \texttt{E}_\lambda \;  
\label{eq:sep}
\end{equation}
\end{definition}

SEP values close to $0$ ($1$) mean that on average the included shared entity has a low (high) popularity. 

{\color{black} The optimization of the selected reasoning paths for SEP can lead to side effects of a type similar to the ones we discussed for LIR. 
It means that it might be possible that all or the majority of the selected reasoning paths focus on a tiny set of very popular shared entities (extreme case: all the selected reasoning paths mention the most popular entity).
This phenomenon could for instance introduce filter bubbles in the space of shared entities mentioned in the explanations and negatively impact on the perceived explanation quality. 
We therefore introduce another metric for monitoring the diversity of the shared entities included in the selected reasoning paths}. Formally:

\begin{definition}
{{\color{black}(Shared Entity Diversity - SED)} Given a user $u$ and the set of reasoning paths $\Tilde{\mathcal L}^k_u$ selected for explaining the recommended products $\tilde{\mathcal P}_u$, with $|\Tilde{\mathcal L}^k_u|=|\tilde{\mathcal P}_u|$, the shared entity diversity over those reasoning paths is:}
\vspace{-2mm}
\begin{equation}
SED(\Tilde{\mathcal L}^k_u) = \frac{|\{e_{k-1} \; | \; \forall l = \{ e_0 \xleftrightarrow[]{r_1} ... \xleftrightarrow[]{r_k} e_k\} \; \in \; \Tilde{\mathcal L}^k_u \}|}{|\Tilde{\mathcal L}^k_u|}
\label{eq:sed}
\end{equation}
\end{definition}

$SED$ values lay in $(0, 1]$, with values close to $0$ ($1$) meaning that the explanations for the recommended list have a low (high) shared entity diversity. 

\subsection{Diversity of the Explanation Path Type}
{\color{black}The third aspect was the diversity of path type (e.g., directed and starred in the movie domain), for a list of explanations.} 

\vspace{1mm}\noindent\textbf{Motivation.} From our literature analysis, we observed that, in psychological science, information diversity is considered a key factor affecting human comprehension and decisions~\cite{AdelmanBQ06}. 
In RS research, diversity is becoming increasingly important, arguing that recommending products by only their predicted relevance increases the risk of producing results that do not satisfy users because the products tend to be too similar to each other~\cite{DBLP:journals/tiis/KaminskasB17, 10.1007/978-3-642-38844-6_16}.  
Considering explanations provided in a recommended list as a whole, a possible conceptualization of diversity is that  the more reasoning path types we present, the better the explanations are perceived. 
For example, in the music domain, we might consider reasoning path types including \texttt{featured} (as in the example path), \texttt{wrote by}, and \texttt{composed by}, and aim at covering them in the provided explanations in a reasonably balanced way. 
Explanation diversity can help countering the dominance of collaborative-based explanations, in the form ``... because a user who listened to your recommended song has also listened to another song you know". This phenomena has been highlighted by prior works (e.g., \cite{10.1145/3404835.3462847}) and confirmed in our exploratory analysis (see the appendix).
We therefore conjecture that the imbalanced representation of the relations in a knowledge graph is one of the main causes. 
This reasoning path type might be deemed as too generic - users receiving the recommendation would not know who the other user is, so they cannot trust them  \cite{Tintarev2007}. 
Moreover, the results of our users assessment showed that $70\%$ of the participants were in favor of the recommended list accompanied by highly diverse reasoning path types. 
Surprisingly, $25.7\%$ of the participants expressed their preference towards a low diversity.
While, $4.3\%$ of the participants declared that this property would not be relevant.

\vspace{1mm}\noindent\textbf{Operationalization.}
Motivated by our findings, we introduced a metric able to quantify how many different types of path accompany the list of recommendations. 
As a proxy of the explanation type, we consider the last relation $r_{k}$ in a certain reasoning path $l^k_{u,p}$ (see def. \ref{def:path-type}).
This metric is computed as the number of unique types of explanations accompanying the recommendations divided by the minimum between (i) the total number of explanations and (ii) the total number of possible relations in $\mathcal G$. Formally: 
 
\begin{definition}
{{\color{black}(Explanation Path Type Diversity - PTD)} Given a user $u$ and the set of reasoning paths $\Tilde{\mathcal L}^k_u$ selected for explaining the recommended products $\tilde{\mathcal P}_u$, with $|\Tilde{\mathcal L}^k_u|=|\tilde{\mathcal P}_u|$, the explanation path type diversity is:}
\vspace{-2mm}
\begin{equation}
PTD(\Tilde{\mathcal L}^k_u) = \frac{|\{r_k \; | \; \forall l = \{ e_0 \xleftrightarrow[]{r_1} ... \xleftrightarrow[]{r_k} e_k\} \in \Tilde{\mathcal L}^k_u \}|}{min(|\Tilde{\mathcal L}^k_u|, \;| \mathcal R |)} 
\label{eq:PTD}
\end{equation}
\end{definition}

PTD values lay in the range $[0, 1]$, with values close to $0$ ($1$) meaning that the recommended list has a low (high) explanation path type diversity. 

Since PTD takes into account the presence of path types in the recommended list but not their representation, {\color{black} the optimization of this metric might lead to cases where certain path types are still over-represented (e.g., those associated to collaborative-based explanations, in the form ``... because a user who listened to your recommended song has also listened to another song you know") and others potentially of higher interest and relevance for a user might appear just once in the recommendations.}
We therefore decided to additionally monitor the concentration of the explanation path types over the selected reasoning paths.
In this way, we can assess whether explanation path types are both diverse and follow a well-balanced representation over the recommendations. 
This additional metric is computed using the Inverse Simpson Index \cite{simpson1949measurement}, measuring the probability of picking two explanations with a different type in the recommended list \cite{FuXGZHGXGSZM20}. Formally:

\begin{definition}
{{\color{black}(Explanation Path Type Concentration - PTC)} Given a user $u$ and the set of reasoning paths $\Tilde{\mathcal L}^k_u$ selected for explaining the recommended products $\tilde{\mathcal P}_u$, with $|\Tilde{\mathcal L}^k_u|=|\tilde{\mathcal P}_u|$, the explanation path type concentration is:}
\begin{equation}
  \begin{aligned}
PTC(\Tilde{\mathcal L}^k_u) = 1 - \frac{\sum_{r \in \mathcal R}^{} \mathcal N(r) \; (\mathcal N(r)-1)}{|\Tilde{\mathcal L}^k_u| \;  (|\Tilde{\mathcal L}^k_u|-1)} 
\\ 
\text{\bf{with}} \; \mathcal N (r) = \{l \; | \; \forall l = \{ e_0 \xleftrightarrow[]{r_1} ... \xleftrightarrow[]{r_k} e_k\} \in \Tilde{\mathcal L}^k_u \} \; \wedge \; r_k = r) \}
\label{eq:PTC}
  \end{aligned}
\end{equation}
\end{definition}

PTC values lay in the range $(0, 1]$, with values close to $0$ meaning that the representation of the explanation path types covered in the recommended list is not well-balanced (one or few types are more represented than others). 

\section{Explanation Property Optimization}\label{sec:algorithm}
{\color{black} Optimizing a recommendation model for the reasoning path metrics formalized in the previous section is the other key task.}
To this end, we propose two classes of approaches. The first class (in-processing) includes approaches that embed reasoning path quality properties in the internal model learning process.
Whereas, the second class (post-processing) covers approaches that re-arrange the recommended lists (and the explanations) returned by the original recommendation model optimized only for recommendation quality. 
These two classes have mutual advantages and disadvantages. 
Post-processing approaches might be limited in their impact, since reordering a small set of recommendations (and explanations) might have a less profound effect than optimizing them during the training process. 
However, post-processing approaches can be applied to any recommendation model and more easily extended to any new metric.

\subsection{In-Processing Optimization}
Our goal is to generate product recommendations accompanied by reasoning paths, as formalized in our KGRE-Rec task (See def.~\ref{def:kbre-problem}), considering both recommendation quality and reasoning path quality. 
We propose to model the problem behind this task as a Markov Decision Process (MDP), which first generates candidate paths between users and products based on a certain similarity measure, and then performs a sampling among candidate paths (see \cite{XianFMMZ19}). 
To solve this problem, we adopted a reinforcement learning (RL) strategy. 
We describe its components, i.e., state initialization, state transition, reward definition, and candidate path sampling, below. 

\vspace{1mm}\noindent\textbf{State Initialization.} Given a path length $k$, the state of the RL agent at step $i$, with $i \leq k$, is defined as a tuple $(u, e_i, h_i)$, where $u \in \mathcal U$ is the user entity, $e_i$ is the entity the RL agent has reached at step $i$, and $h_i = \{r_1, \ldots, e_{i - 1}, r_i\}$ is the historical trace path followed by the RL agent until step $i$. Given a user $u \in \mathcal U$, the initial state is $(u, u, \emptyset)$ and the terminal state is $(u, e_k, h_k)$. In the initial state, users were uniformly sampled. We constrained the path length to $k=3$, proved to lead to the best performance in prior work \cite{ripple-net/10.1145/3269206.3271739}, and forced the entity $e_k$ to be a product (to be recommended). 

\vspace{1mm}\noindent\textbf{State Transition}. The complete action space $\mathcal A_i$ of state $i$ is defined as all possible outgoing edges of entity $e_i$ in the $\mathcal G$, excluding any entities and relations in $h_i$. Formally, $\mathcal A_i = \{(r, e) \; | \; (e_i, r, e) \in \mathcal G, e \notin h_i, r \notin h_i\}$. Since the out-degree followed a long-tail distribution, some nodes had much higher out-degrees compared with the others. Thus, we introduce an action pruning strategy that keeps edges conditioned on the user based on a scoring function $\Psi : (\mathcal U, \mathcal E, \mathcal R) \rightarrow \mathbb R$. Then, the pruned action space of state $i$ for user $u$ is defined as $\Tilde{\mathcal A}_i(u) = \{(r, e) \; | \; rank(\Psi_i( \, (u, e_i, h_i), \, (r, e) \, )) \leq Z_i, (r, e) \in \mathcal A_i\}$, where $Z \in \mathbb N$ is the maximum action space size at step $i$.
The scoring function $\Psi$ is typically the dot product between the head entity $e_{i}$ and the tail entity $e$ of state $i$, both represented via translational embeddings \cite{BordesUGWY13}. 
Therefore, during pruning, reasoning paths that might be relevant according to the notions of reasoning path quality might end up being excluded.
To counter this, we extended the original multi-hop scoring function with a second term associated to the reasoning path quality, according to the hop that has been reached. 
Formally, given a state $(u, e_i, h_i)$ and an action $(r, e) \in \mathcal A_i$:  

\begin{equation}
\begin{aligned}
\Psi_i( \, (u, e_i, h_i), \, (r, e) \, ) = \;  (1 - \alpha ) \; \Tilde{\mathcal F}(e_i, e)  \; + \\ \alpha \sum_{c \; \in \; \mathcal C} \; c(\{[u] \cdot h_i \cdot [e_i, r, e] \}) \\
\textbf{with} \; \Tilde{\mathcal F}(e_i, e) = <\textbf{e}_i, \textbf{e}> \; + \;  b_{e}
\label{eq:psi-transit}
\end{aligned}
\end{equation}

where $< \dots >$ is the dot product operation, $\cdot$ is the list concatenation operation, $\bf{e_i}, \bf{e} \in \mathbb R^d$ are the $d$-dimensional embeddings of the entities $e_i, e \in \mathcal E$, $b_{e} \in \mathbb R$ is the bias for entity $e$, $\alpha \in [0, 1]$ is the weight assigned to the explanation term (optimized through grid search in our study), and $\mathcal C \subset \mathcal C^{EQ} = \{ LIR, SEP, PTD\}$ is a subset of reasoning path quality metrics to optimize.
{\color{black} At step $i=0$, the entity $e$ refers to a product already experienced by user $u$.
LIR would be the only reasoning path quality metric that can be computed ($\mathcal C = \{ LIR \}$). 
Given that $k=3$, the entity $e$ at step $i=1$ refers to the entity shared between the already experienced product and the product to be recommended and also uniquely determines the type of path ($\mathcal C = \{ LIR, SEP, PTD \}$).}
Since $PTD$ is a property of a set of paths, we gave a higher score for this metric when the type of the current path was not already seen for user $u$ during learning.  

Finally, the action $(r, e) \in \Tilde{\mathcal A}_i(u)$ with the maximum $\Psi_i( \, (u, e_i, h_i), \, (r, e) \, )$ was selected to transit to the next state, during the learning process.

\vspace{1mm}\noindent\textbf{Reward Definition.} Based on our MDP formulation, our goal is to learn a stochastic policy $\pi$ that maximizes the expected cumulative reward for any initial user $u$. To solve this task, we adopted the same REINFORCE strategy proposed by \cite{XianFMMZ19}. However, their original cumulative reward function represented a proxy of recommendation utility only (and hence encodes recommendation quality - $\mathcal C^{RQ}$) and does not consider any notion of reasoning path quality. To counter this, using the same notation of Eq. \ref{eq:psi-transit}, we computed the reward function $\varphi$ on the final state $(u, e_k, h_k)$ as:

\vspace{-5mm}
\begin{equation}
\varphi(u, e_k, h_k) = \;  (1 - \alpha ) \; \Tilde{\mathcal F}(u, e_k)  \; + \; \alpha \; \sum_{c \; \in \; \mathcal C} c(\{ [u] \cdot h_k \cdot [e_k] \})
\end{equation}

where $\bf{u}, \bf{e_k} \in \mathbb R^d$ are the $d$-dimensional embeddings representations of the user entity $u$ and the recommended product $e_k$ respectively, and $\mathcal C = \mathcal C^{EQ} = \{ LIR, SEP, PTD\}$ is the set of path quality metrics to optimize. 

\vspace{1mm}\noindent\textbf{Candidate Path Sampling.} 
Given the learnt stochastic policy $\pi$ and a user $u$, the final step is to infer the set of $n$ products $\Tilde{\mathcal P}^n_u$ to be recommended and the set of reasoning paths $\tilde{\mathcal L}^k_{u}$ used for generating their textual explanations. 

To extract the candidate paths (and the recommended products) for each user $u$, we applied a beam search guided by the action probability encoded in the policy $\pi$ \cite{XianFMMZ19} (see Algorithm \ref{alg:guided-path-reasoning}). 
This procedure takes as an input a given user $u$, the learnt policy network $\pi$, the path length $k$ ($k=3$ in our study), and the sampling sizes denoted by $Z_1, \ldots, Z_k \in \mathbb N$. 
As an output, it yields a set of candidate user-product paths $\mathcal{L}^k_{u}$, with each path associated to (i) the generative probability $\mathcal s^k$ in the set $\Tilde{\mathcal S}$ and (ii) the relevance score $\Tilde{\mathcal F}(u, p) \in \Phi^k$ of that product $p$ for $u$. 

Within the set of candidate paths $\mathcal{L}^k_u$, there may exist multiple paths from user $u$ to the same product to recommend $p$. 
Thus, for each product $p \in \{e_k \; | \; \forall l = \{ e_0 \xleftrightarrow[]{r_1} ... \xleftrightarrow[]{r_k} e_k\} \in \mathcal{L}^k_u \}$, we selected the path $\Tilde{l}^k_{u,p} \in \mathcal {L}^k_{u,p}$ with the highest generative probability based on $\Tilde{\mathcal S}$. Being the model optimized also for reasoning path quality, such probability score encoded information about it as well.
Finally, we ranked the selected paths based on the relevance $\Tilde{\mathcal F}(u, p) \in \Phi^k$ and recommended the products $\Tilde{\mathcal P}^n_u$ to $u$.

\begin{algorithm}[!ht]
{\color{black}
\caption{Candidate path sampling constrained to path quality}\label{alg:guided-path-reasoning}
\KwData{knowledge graph $\mathcal G$, user $u$, learnt policy $\pi$, path length $k$, action sampling sizes ${Z_1, \ldots, Z_k}$, relevance function $\Tilde{\mathcal F}$}
\KwResult{candidate user-product path set $\mathcal L^k_u$, probability set $\Tilde{\mathcal S^k}$, relevance set $\Phi$}
$\mathcal L^0_{u} \gets \{\{u\}\}$, $\mathcal S^0 \gets \{1\}$ ; \\
$state_0 \gets (u, u, \emptyset)$ \tcp*[l]{\footnotesize State initialization}
\For{$i \gets 1$ to $k$}{ 
$\mathcal L^i_{u} \gets \emptyset$, $\mathcal S^i \gets \emptyset$ ;\\
\ForAll{$l = \{\dots, e_{i-1}\} \in \mathcal{L}_u^{i-1}, \mathcal s \in \mathcal S^{i-1}$}{
\tcc{\footnotesize Actions candidate for state transition (with pruning) }
$\mathcal A_i \gets \{a = (r_i, e_i) \; | \; (e_{i-1}, r_i, e_i) \in \mathcal G, e_i \notin l, r_i \notin l\}$ ;\\
$\Tilde{\mathcal A}_i \gets \{a \in \mathcal A_i \; | \; rank(\pi(state_{i-1}, a)) \leq Z_i\}$ ;\\
\ForAll{$a=(r_i, e_i) \in \Tilde{\mathcal A}_i$}{
$state_{i} \gets (u, e_i, l \cup \{r_i\})$ \tcp*[l]{\footnotesize State update}
$\mathcal L^i_{u}$ $\gets$ $\mathcal L^i_{u}$ $\cup$ $\{ l \cup \{r_i, e_i\} \}$ \tcp*[l]{\footnotesize Add extended path}
$\mathcal S^i$ $\gets$ $\mathcal S^i$ $\cup$ $\{$$\mathcal{s} \; \pi(state_i, a) \}$ \tcp*[l]{\footnotesize Add probability score}
}
}
}
\tcc{\footnotesize Path, probability, relevance for paths ending with a product}
$\mathcal L^k_{u} \gets \{l \, | \, \forall l = \{e_0, \dots, e_k\} \in \mathcal L^k_{u} \wedge e_k \in \mathcal P$\} ; \\
$\Tilde{\mathcal S}^k \gets \{\mathcal S^k_l \, | \, \forall l \in \Tilde{\mathcal L}^k_{u}\}$ ; \\
$\Phi^k \gets \{\Tilde{\mathcal F}_{u, e_k} \, | \, \forall l = \{u, \dots, e_k\} \in \mathcal L^k_{u} \}$ ; \\
\Return $\mathcal L^k_{u}$, $\Tilde{\mathcal S}^k$, $\Phi^k$\\
}
\end{algorithm}

\subsection{Post-Processing Optimization}
Compared to an in-processing optimization, post-processing approaches perform a re-ranking of both products to recommend and their corresponding reasoning paths, to optimize certain reasoning path quality metrics. 
The input of such step are the recommended products and their selected reasoning paths, originally returned by any pre-trained model that solves the KGRE-Rec task. 
For the implementation, we capitalized on a \emph{maximum marginal relevance} approach, with the reasoning path metric(s) as support metric(s). 

{\color{black} For each position $i \leq n$ of the recommended list, for each candidate path $l \in \mathcal{L}^k_u$, we computed a weighted sum between (i) the relevance of the product $p$ associated to the path $l$ for the user $u$ ($\Phi^k_l = \Tilde{\mathcal F}(u, p)$) and (ii) the extent to which the recommended list re-ranked so far would increase the target metric, if we had included that product $p$ with a specific path $l$ in the recommendations at position $i$.
Once we computed this weighted score for all paths, we found the path $\Tilde{\mathcal L}^{k,i}_{u} = \{u, \dots, e_k\}$ that achieves the highest weighted score, recommended the last entity $\Tilde{\mathcal P}^i_u = e_k$ (a product) to user $u$ at position $i$, and generated its textual explanation based on the path $\Tilde{\mathcal L}^{k,i}_{u}$. 
This procedure is repeated until position $n$ (see Algorithm \ref{alg:post-proc}).
Formally, the product $\Tilde{\mathcal P}^i_u$ and the selected path $\Tilde{\mathcal L}^{k,i}_{u}$ are determined as:} 

\vspace{-4mm}

\begin{equation}\label{eq:opt_prob}
\begin{aligned}
    (\Tilde{\mathcal P}^i_u, \Tilde{\mathcal L}^{k,i}_{u}) = \mathop{\text{argmax}}_{p \; \in \; \mathcal P \setminus \Tilde{\mathcal P}_u^{i-1}, \; l \; \in \; \mathcal L^k_u \setminus \Tilde{\mathcal L}^{k,i-1}_{u}} \, \mathcal Q(l=\{u, \dots, p\}, \Phi^k_l, \mathcal C) \\ 
    \textbf{with} \; \mathcal Q(l, \Phi^k_l, \mathcal C) = (1-\alpha) \; \Phi^k_l + \alpha \,  \mathcal \sum_{c \; \in \; \mathcal C} c(l)
\end{aligned}
\end{equation}

\noindent where $\alpha \in [0,1]$ is a parameter that expresses the trade-off between relevance and the target (set of) reasoning path metrics, and $\mathcal C$ is the set of reasoning path quality metrics (either LIR, SEP, PTD, or any combination) to optimize for.
With $\alpha=0$, we yield the output of the original model, not considering reasoning path quality (viceversa for $\alpha=1$).
This greedy approach fits with the real world, where users may see only the first $n$ recommendations, and the rest becomes visible after scrolling\footnote{
For completeness, we refer the reader to our repository for a detailed description of the computational complexity each of the proposed methods has. 
}. 

\begin{algorithm}[!t]
{\color{black}
\caption{Candidate path sampling constrained to path quality}\label{alg:post-proc}
\KwData{candidate paths set $\mathcal L^k_{u}$, relevance set $\Phi^k$, re-ranking function $\mathcal Q$, path quality functions $\mathcal C$, recommended list size $n$}
\KwResult{recommended products list $\Tilde{\mathcal P}^n_u$, reasoning paths list $\Tilde{\mathcal L}^k_{u}$}

$\Tilde{\mathcal P}^0_u \gets \emptyset, \Tilde{\mathcal L}^{k,0}_{u} \gets \emptyset$ ;\\
$\hat{\mathcal{P}} \gets \{e_k \; | \; \forall l = \{u, \dots, e_k\} \in \mathcal L^k_u \land e_k \in \mathcal P\}$ \tcp*[l]{\footnotesize Reachable products}
\ForAll{$i \gets 1$ to $n$}{
    $\Tilde{\mathcal L}^{k,i}_{u} \gets \Bigl\{ \underset{l  = \{u, \dots, e_k\} \; \in \; \mathcal{L}^k_u}{\mathrm{argmax}} \, \mathcal{Q}(l, \Phi^k_l, \mathcal C) \Bigl\}$ ;\\
    $\Tilde{\mathcal P}^i_u \gets \{ e_k \} $ ; \\
    $\mathcal{L}^k_u \gets \mathcal{L}^k_u \; \setminus \; \Tilde{\mathcal L}_u^{k,i}$ \tcp*[l]{\footnotesize Remove the identified path}
    $\hat{\mathcal{P}} \gets \hat{\mathcal{P}} \; \setminus \; \Tilde{\mathcal P}_u^{i}$ \tcp*[l]{\footnotesize Remove the recommended product}
}
\Return $\Tilde{\mathcal P}^n_u, \Tilde{\mathcal L}^{k,n}_{u}$
}
\end{algorithm}

\section{Experimental Evaluation}\label{sec:experiments}
In this section, we aim to evaluate the proposed suite of approaches, by answering to the following key research questions:

\vspace{-0.5mm}
\begin{enumerate}[label={\textbf{RQ\protect\threedigits{\theenumi}}}, leftmargin=*]
    \item How does the recommendation utility achieved after applying our approaches compare to that achieved by other state-of-the-art models?
    \item Which type of optimization (in-processing, post-processing, their combination) leads to the highest reasoning path quality?
    \item {\color{black} How do our approaches affect the final recommendations and explanations provided to users?}
\end{enumerate}

\subsection{Experimental Setup}\label{sub:setup}

\vspace{1mm}\noindent\textbf{Data Sets.} 
We conducted experiments on three data sets, namely MovieLens1M (ML1M) \cite{ml1m}, LastFM-1B (LASTFM) \cite{lastfm-dataset}, and Amazon-Cellphones (CELL) \cite{ni-etal-2019-justifying}. 
They are all public and vary in domain, extensiveness, and sparsity (see Table \ref{tab:data-stats}). 
For ML1M, we adopted the KG generated in \cite{CaoWHHC19}, whereas we adopted the KG generated in \cite{Wang00LC19} for LASTFM. 
For CELL, the KG was built from contextual user reviews, each including related products, the product subcategory, and the review text, as done by \cite{10.1145/3132847.3132892}.
For all data sets, we discarded (i) products (and their interactions) absent in the KG as an entity and (ii) relations occurring less than 200 times.

\begin{table}[!t]
  \caption{Statistics of the pre-processed data sets used in this study.}
  \label{tab:data-stats}
  \centering
\resizebox{1\linewidth}{!}{
  \begin{tabular}{lrrrrr}
    \newline & \newline & \textbf{ML1M} & \textbf{LASTFM} & \textbf{CELL} \\
    \hline
    \toprule
    User-Product & \# Users & 6,040 & 15,773  & 27,879 \\
    Information & \# Products & 3,226 & 47,981 & 10,429 \\
    \newline & \# Interactions & 1,000,209 & 3,955,598 & 194,439\\
    \hline
    Knowledge & \# Entities & 16,899 & 114,255 & 425,264 \\
    Graph & \# Relations & 1,830,766 & 8,501,868 & 477,409\\
     & \# Relation Types & 10 & 9 & 7\\
    \bottomrule
  \end{tabular}
  }
\vspace{4mm}
\end{table}

\vspace{1mm}\noindent\textbf{Data Preparation.} 
For each data set, we first sorted the interactions of each user chronologically. 
We then performed a training-validation-test split with the 70\% oldest interactions in the training set, the subsequent 10\% in the validation set (adopted for hyper-parameter fine tuning), and the 20\% most recent interactions in the test set. 
We assumed that products already seen by the user were not recommended another time. 
The same pre-processed data sets were used to train, optimize, and test each benchmarked model.

\vspace{1mm}\noindent\textbf{Evaluation Metrics.} 
We monitored recommendation quality and reasoning path quality on top-10 recommended lists ($n=10$), a well-known recommendation cut-off. 
For the first perspective, we measured the Normalized Discounted Cumulative Gain (NDCG) \cite{WangWLHL13}, using binary relevance scores and a base-2 logarithm decay, and the Mean Reciprocal Rank (MRR).
For the second perspective, we measured the Linking Interaction Recency (LIR, Eq. \ref{eq:lir}), the Shared Entity Popularity (SEP, Eq. \ref{eq:sep}), and the Explanation Path Type Diversity (PTD, Eq. \ref{eq:PTD}). 
In addition, we monitored the Linking Interaction Diversity (LID, Eq. \ref{eq:lid}), the Shared Entity Diversity (SED, Eq. \ref{eq:sed}), and the Explanation Path Type Concentration (PTC, Eq. \ref{eq:PTC}).

\vspace{1mm}\noindent\textbf{Benchmarked Models.} 
The considered baselines\footnote{The detailed description of each baseline and the selected fine-tuned
hyper-parameters are listed in the README of our \href{https://tinyurl.com/bdbfzr4n}{repository}.}
included factorization models (BPR \cite{DBLP:conf/uai/RendleFGS09}, FM \cite{10.1145/2009916.2010002}, NFM \cite{DBLP:conf/uai/RendleFGS09}), three knowledge-aware models based on regularization (CKE \cite{CKE10.1145/2939672.2939673}, CFKG \cite{AiACZ18}, KGAT \cite{Wang00LC19}), and one knowledge-aware model based on reasoning paths (PGPR \cite{XianFMMZ19}). 
Our optimization approaches could be applied only to models based on reasoning paths (i.e., PGPR).
In a single-metric optimization scenario\footnote{For conciseness, we did not cover a joint optimization of multiple reasoning path quality metrics. We leave it as a future work.}, we denote as $\{R|P|D\}$-(P)PGPR the results after post-processing only, as $\{R|P|D\}$-(I)PGPR the results after in-processing only, and as $\{R|P|D\}$-(IP)PGPR the results combining both, based on the optimized reasoning path quality metric: recency (R : LIR), popularity (P : SEP), diversity (D : PTD).   

\subsection{RQ1: Impact on Recommendation Utility}\label{sec:performance-comparison-rq}
In a first analysis, we compared the utility of recommendations obtained after applying our approaches and of those obtained by the baseline models\footnote{
For our approaches, the values of $\alpha$ were selected assuming that the platform owners are willing to lose $10\%$ of NDCG at most to increase as much as possible reasoning path quality. While scientists bring forth the discussion about metrics and design models optimized for them, it is up to stakeholders to select the trade-off most suitable.}.
With this in mind, Table~\ref{tab:baselines-results} reports the NDCG and MRR obtained by each model under the considered settings. The patterns for both metrics were similar and, therefore, we will focus only on NDCG below.

\begin{table}[!t]
\centering
  \caption{Recommendation utility (NDCG@10, MRR@10).}
  \label{tab:baselines-results}
  \resizebox{1\linewidth}{!}{
\begin{tabular}{rrrrrrr|rrrrrrr}
 \toprule
\multicolumn{7}{c}{\textbf{Baselines $\uparrow$}}                                                            &            & \multicolumn{6}{c}{\textbf{Ours $\uparrow$}}                                                         \\
Model & \multicolumn{2}{c}{ML1M} & \multicolumn{2}{c}{LASTFM} & \multicolumn{2}{c}{CELL} & Model      & \multicolumn{2}{c}{ML1M} & \multicolumn{2}{c}{LASTFM} & \multicolumn{2}{c}{CELL} \\
      & NDCG        & MRR        & NDCG         & MRR         & NDCG        & MRR        &            & NDCG        & MRR        & NDCG         & MRR         & NDCG        & MRR        \\
          \midrule

BPR   & \textbf{0.33}        & 0.23          & 0.13         & \underline{0.09}           & 0.05        & \underline{0.03}          & R-(I)PGPR  & \textbf{0.33}        & \textbf{0.25}          & 0.13         & \underline{0.09}           & \underline{0.06}        & 0.02          \\
FM    & \underline{0.32}        & 0.21          & \textbf{0.15}         & \underline{0.09}           & 0.02        & 0.01          & R-(P)PGPR  & \textbf{0.33}        & \textbf{0.25}          & \underline{0.14}          & 0.08           & \underline{0.06}        & \textbf{0.04}          \\
NFM   & \underline{0.32}        & 0.21          & \underline{0.14}         & \underline{0.09}           & 0.02        & 0.01          & R-(IP)PGPR & \textbf{0.33}        & \textbf{0.25}          & \underline{0.14}          & \underline{0.09}           & \underline{0.06}        &   \textbf{0.04}        \\
CKE   & \textbf{0.33}        & 0.22         & \underline{0.14}          & \underline{0.09}           & 0.04        & \underline{0.03}          & P-(I)PGPR  & \textbf{0.33}        & \underline{0.24}          & 0.12         & 0.08          & \textbf{0.07}        & \underline{0.03}          \\
CFKG  & 0.27        & 0.17          & 0.08         & 0.05           & 0.00        & 0.00          & P-(P)PGPR  & \underline{0.32}        & \underline{0.24}          & \underline{0.14}          & 0.08           & \underline{0.06}        & \underline{0.03}          \\
KGAT  & \textbf{0.33}        & \underline{0.24}          & \textbf{0.15}         & \underline{0.09}           & 0.02        & 0.01          & P-(IP)PGPR & \underline{0.32}        & 0.23          & 0.12         & 0.08           & \underline{0.06}        & \underline{0.03}          \\
PGPR  & \textbf{0.33}        & \underline{0.24}          & \underline{0.14}         & \underline{0.09}           & \textbf{0.07}        & \textbf{0.04}          & D-(I)PGPR  & \textbf{0.33}        & \textbf{0.25}          & \underline{0.14}          & \textbf{0.10}           &  \textbf{0.07}        & \textbf{0.04}          \\
      &             &            &              &             &             &            & D-(P)PGPR  & 0.31        & \underline{0.24}          & \textbf{0.15}         & \underline{0.09}           & \underline{0.06}        & \textbf{0.04}          \\
      &             &            &              &             &             &            & D-(IP)PGPR & \underline{0.32}        & \underline{0.24}          & \underline{0.14}         & \underline{0.09}           & \underline{0.06}        & \textbf{0.04}  \\    
\bottomrule
  \multicolumn{14}{l}{For each data set: best result in \textbf{bold}, second-best result \underline{underlined}.}\tabularnewline
\end{tabular}  
}
\vspace{4mm}
\end{table}

Except for CFKG, the considered baselines achieved an overall similar NDCG ranging in [0.32, 0.33] (ML1M) and  [0.13, 0.15] (LASTFM). In CELL, the NDCG was generally low, ranging in [0.02, 0.05], except for PGPR (0.07). This result might be caused by the high sparsity and low test set size of the CELL set. Interestingly, in this data set, the gap in performance between traditional recommendation models and PGPR was higher. The main reason could be that PGPR exploited information from the review linked to a certain user interaction for generating better recommendations.

Compared to the baselines, all our approaches did not substantially affect NDCG (loss $\leq 0.01$ NDCG points). 
In ML1M, five out of nine cases resulted in the same NDCG achieved by the best baseline model (0.33), three out of five cases showed a decrease of $0.01$ NDCG points, and just in one case the decrease was of $0.02$ NDCG points (D-(P)PGPR). 
Similarly, in CELL, two out of nine cases resulted in a NDCG comparable to that of the best baseline model ($0.07$), whereas the other seven cases were characterized by a decrease of just $0.01$ NDCG points. 
The NDCG of our approaches in LASTFM was more evidently worse than that of baseline models, compared to the other data sets. 
Just one case, D-(P)PGPR resulted in an NDCG equal to that of the best baseline model, whereas the other cases showed a decrease in NDCG between $0.01$ and $0.03$ NDCG points.  
We conjecture that this difference in LASTFM might be caused by the peculiar nature of listening interactions.      

Considering in- and post-processing approaches, the former optimization strategy led to the highest NDCG in ML1M (0.33, regardless of the optimized property) and CELL (0.07 in case of SEP and PTD optimization, 0.06 for LIR optimization).
The latter optimization strategy resulted in a higher NDCG in LASTFM (0.15 under a PTD optimization, 0.14 for both LIR and SEP). 
Comparing reasoning path quality metrics, optimizing for LIR (SEP and PTD) generally led to the highest NDCG in ML1M (LASTFM and CELL). 
We conjecture that the reason behind this difference across data sets comes from the peculiar temporal patterns of the user interactions in ML1M, emerged from our exploratory analysis (see the appendix).     

\vspace{-1mm}
\hlbox{Observation 1}{
Optimizing for reasoning path quality through our approaches led to state-of-the-art NDCG. 
The measured NDCG was equal or at most two points lower than that of non-(path-)explainable baselines, on all data sets. 
It emerged that accounting for user-level reasoning path quality often does not lead to a loss (when observed, it is negligible) in recommendation utility.}

\subsection{RQ2: Impact on Reasoning Path Quality}\label{sec:exp-rq}
In a second analysis, we investigated the impact of (a combination of) our in- and post-processing approaches on reasoning path quality, compared the that of the original PGPR model. 
Indeed, PGPR represented the only model able to generate reasoning paths, among the considered baselines. 
Table~\ref{tab:results-exp-metrics} collects all the reasoning path quality metrics introduced in our study, namely LIR (LID), SEP (SED), and PTD (PTC), for each model. 
Since the patterns observed in LASTFM and CELL were similar to those observed in ML1M, we describe only the latter ones in detail for conciseness. 

\begin{table*}[!t]
\centering
  \caption{Reasoning path quality (LIR, LID; SEP, SED; PTD, PTC).}
    \label{tab:results-exp-metrics}
    \resizebox{1.0\linewidth}{!}{
    \begin{tabular}{r|rrrrrr|rrrrrr|rrrrrr}
    \toprule
     & \multicolumn{6}{c}{\textbf{ML1M}} & \multicolumn{6}{c}{\textbf{LASTFM}} & \multicolumn{6}{c}{\textbf{CELL}}\\ 
     & LIR $\uparrow$ & LID $\uparrow$ & SEP $\uparrow$ & SED $\uparrow$ & PTD $\uparrow$ & PTC $\uparrow$ &
       LIR $\uparrow$ & LID $\uparrow$ & SEP $\uparrow$ & SED $\uparrow$ & PTD $\uparrow$ & PTC $\uparrow$ &
       LIR $\uparrow$ & LID $\uparrow$ & SEP $\uparrow$ & SED $\uparrow$ & PTD $\uparrow$ & PTC $\uparrow$\\
    \midrule
    PGPR & 0.43 & 0.80 & 0.24 & 0.98 & 0.13 & 0.09 &
           0.46 & 0.60 & 0.38 & 0.99 & 0.13 & 0.04 &
           0.21 & 0.69 & 0.57 & 0.99 & 0.30 & 0.46\\
    \hline
    R-(I)PGPR & 0.75 & \bf{0.69} & 0.29 & 0.99 & 0.14 & 0.12 &
              0.66 & \bf{0.63} & 0.41 & 0.99 & 0.16  & 0.09 &
              \underline{0.46} & \bf{0.75} & 0.52 & 0.99 & 0.19 & 0.12 \\
    R-(P)PGPR & 
    \underline{0.89} & 0.42 & 0.40 & 0.99 & 0.14 & 0.12 &
    \underline{0.93} & 0.43 & 0.42 & 0.99 & 0.14 & 0.06 &
    0.33 & \underline{0.57} & 0.81 & 0.99 & 0.36 & 0.45 \\ 
    R-(IP)PGPR & \bf{0.92} & \underline{0.46} & 0.36 & 0.99 & 0.14 & 0.12 &
                \bf{0.93} & \underline{0.42} & 0.45 & 0.99 & 0.15 & 0.07 &
                \bf{0.86} & 0.51 & 0.52 & 0.99 & 0.19 & 0.11 \\
    \hline
    P-(I)PGPR& 0.44 & 0.60 & 0.59 & \bf{0.98} & 0.15 & 0.13 &
             0.55 &  0.71 & \underline{0.67} & \bf{0.99} & 0.17 & 0.13 &
             0.29 &  0.72 & 0.70 & 0.99 & 0.29 & 0.45\\
    P-(P)PGPR & 0.48 & 0.79 & \underline{0.65} & \underline{0.96} & 0.19 & 0.23 &
             0.63 & 0.74 & 0.60 & 0.98 & 0.18 & 0.15 &
             0.33 & 0.57 & 0.81 & 0.99 & 0.36 & 0.45\\ 
    P-(IP)PGPR & 0.43 & 0.78 & \bf{0.78} & 0.96 & 0.16 & 0.15 &
                0.55 & 0.70 & \bf{0.72} & \underline{0.98} & 0.18 & 0.15 &
                0.28 & 0.62 & 0.85 & 0.99 & 0.29 & 0.42\\
    \hline
    D-(I)PGPR& 0.42 & 0.81 & 0.30 & 0.98 & 0.18 & 0.22 &
             0.55 & 0.70 & 0.38 & 0.99 & 0.14 & 0.06 &
             0.34 & 0.64 & 0.80 & 1.00 & 0.20 & 0.04 \\
    D-(P)PGPR & 0.49 & 0.75 & 0.33 & 0.99 & \underline{0.40} & \underline{0.46} &
             0.55 &  0.70 & 0.46 & 0.99 & \underline{0.41} & \underline{0.48} &
             0.24 & 0.68 & 0.41 & 0.99 & \underline{0.47} & \underline{0.62}\\
    D-(IP)PGPR & 0.42 & 0.72 & 0.41 & 0.99 & \bf{0.56} & \bf{0.73} &
                0.54 & 0.70 & 0.48 & 0.99 & \bf{0.50} & \bf{0.59} &
                0.54 & 0.71 & 0.76 & 1.00 & \bf{0.55} & \bf{0.68} \\
    \bottomrule
    \multicolumn{18}{l}{For each category: best result of the optimized metric and its complementary metric in \textbf{bold}, second-best result \underline{underlined}.}\tabularnewline
    \end{tabular}
    }
\vspace{4mm}
\end{table*}

In ML1M, an in-processing optimization already led to a reasoning path quality score higher than that obtained by the original PGPR, respectively for each optimized metric LIR, SEP, and PTD (R-(I)PGPR: 74.4\% LIR and -13.7\% LID; P-(I)PGPR: 145.8\% SEP and 1.02\% SED; D-(I)PGPR: 38.46\% PTD and 144.4\% PTC). 
Each in-processing setting reported gains on the other two reasoning path quality metrics not directly optimized.
This gain was generally higher while optimizing for PTD (-2.32\% LIR, 1.25\% LID; 25\% SEP, 0\% SED) than SEP (2.32\% LIR, -1.25\% LID; 15.3\% PTD, 44.4\% PTC) and LIR (20.8\% SEP, 1.02\% SED; 7.69\% PTD, 33.3\% PTC). 

Similarly, the path quality score obtained via our post-processing was always higher than that of the original PGPR (R-(P)PGPR: 106.9\% LIR and -47.5\% LID; P-(P)PGPR: 174.1\% SEP 1.04\% SED; D-(P)PGPR: 207.6\% PTD and 411.1\% PTC). 
The gain on (non-optimized) path metrics tended to be higher than that observed under in-processing.

Combining in- and post-processing approaches led to the largest gain in reasoning path quality, compared to that of the original PGPR (R-(IP)PGPR: 113.95\% LIR and -42.5\% LID; P-(IP)PGPR: 225\% SEP and 0\% SED; D-(IP)PGPR: 330.7\% PTD and 711.1\% PTC).
Optimizing for either LIR, SEP, or PTD generally resulted in a decrease in LID, SED or PTC respectively, especially under post-processing\footnote{
We refer the reader to our source code repository for additional analyses regarding how the characteristics of the selected reasoning paths were affected by our optimization approaches. 
}. 

\hlbox{Observation 2}{
Compared to PGPR, our in- and post-processing optimization approaches showed a substantially higher reasoning path quality based on the proposed metrics, on all data sets. 
Higher gains were observed for PTD than for the other properties. 
There were gains also on reasoning path metrics not directly optimized, highlighting a positive interdependence across metrics according to the domain.}

{\color{black}
\subsection{RQ3: Impact on Recommendations and Explanations}
In a third analysis, we were interested in understanding how the proposed in- and post-processing approaches concretely changed the top-$n$  recommended lists and their accompanying explanations. 
To this end, we provide a use case for each reasoning path quality perspective, showing the results generated for an example user by (i) the original PGPR model, (ii) our in-processing approach only, (iii) our post-processing approach only, and (iv) the combination of both in- and post-processing. 
For each use case, Fig \ref{fig:qualitative-analysis} reports the explanation template (see def. ~\ref{def:reasoning-path-template}), how the textual explanation can be generated from the selected reasoning path, and the specific part of the explanation influenced by the model output (LIR, SEP, PTD).

\begin{figure}[b!]
\centering
\subfloat[Recency (LIR) with R-(P)PGPR, R-(I)PGPR and R-(IP)PGPR.]{
    \includegraphics[width=0.5\textwidth,height=0.20\textheight]{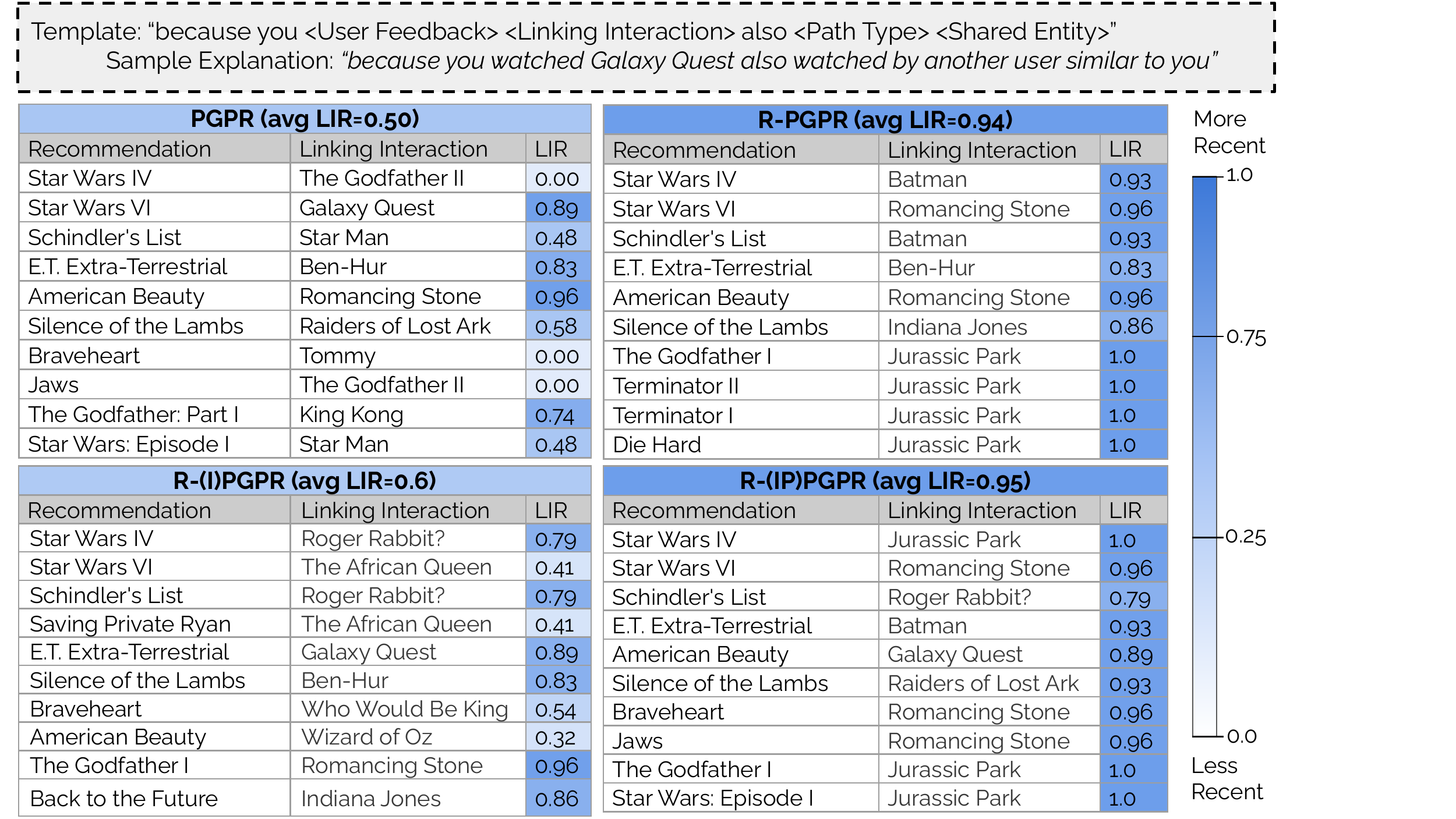}
    \label{fig:case-study-recency}}
 \hfill
\subfloat[Popularity (SEP) with P-(P)PGPR, P-(I)PGPR and P-(IP)PGPR.]{
    \includegraphics[width=0.5\textwidth,height=0.20\textheight]{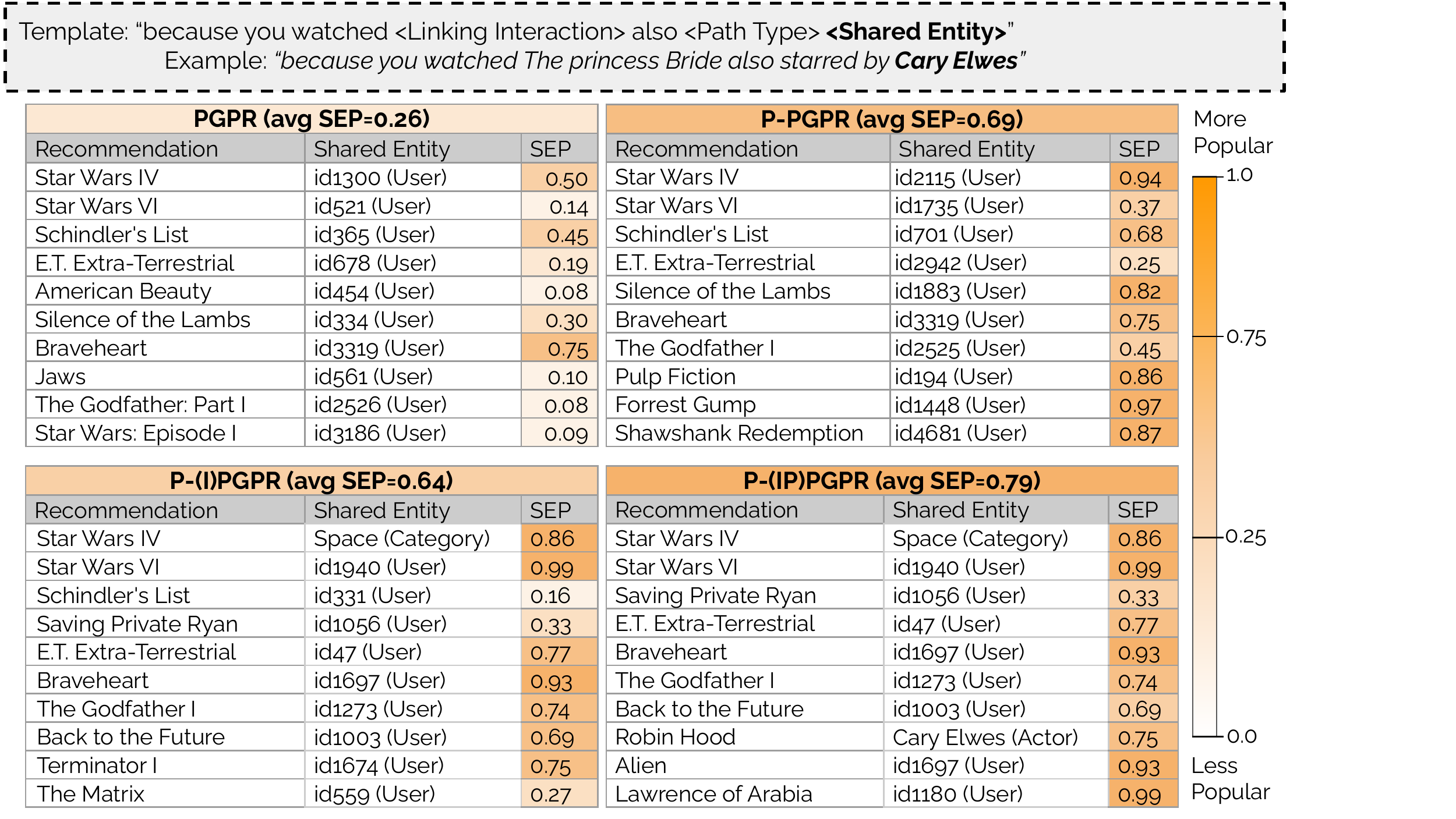}    \label{fig:case-study-popularity}} 
 \hfill
\subfloat[Diversity (PTD) with D-(P)PGPR, D-(I)PGPR and D-(IP)PGPR.]{
    \includegraphics[width=0.5\textwidth,height=0.20\textheight]{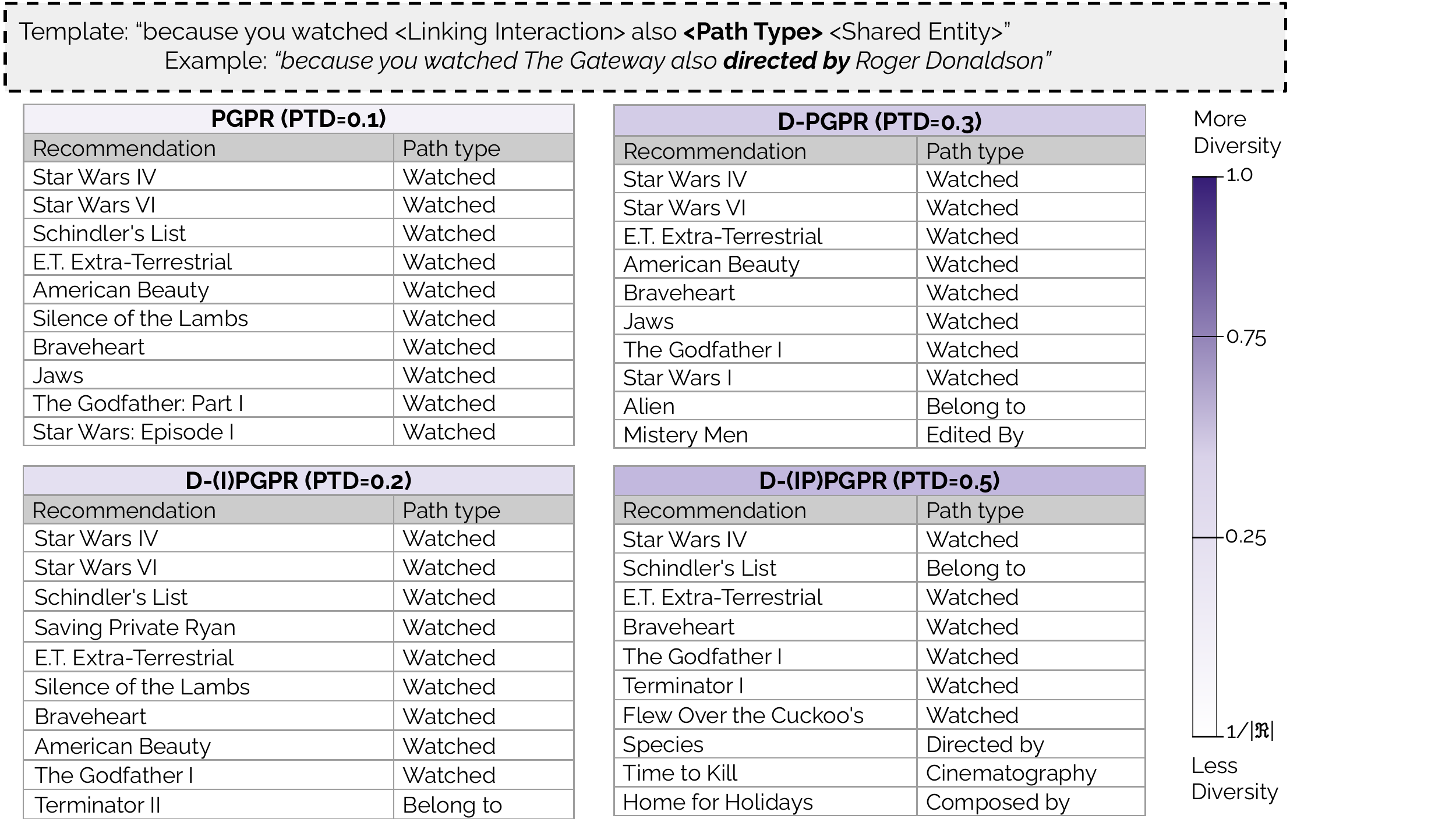}
    \label{fig:case-study-diversity}} 
	
\caption{Impact on the recommendations and explanations for a random user in ML1M.}
\label{fig:qualitative-analysis}

\end{figure}

\vspace{1mm}\noindent\textbf{Recency.} 
Figure~\ref{fig:case-study-recency} shows that our approaches led to textual explanations including more recent linking interactions than the original PGPR model. 
In this specific example, the difference in LIR between the original and the in-processing approaches ($0.10$) and between the post-processing and the combination ($0.01$) was small.
It follows that the main contribution to the increase in LIR was provided by the post-processing optimization. 
However, it should be noted that, these differences increased when we considered the entire user population. 
With this example, however, we were also able to highlight the (negative) side effect that may often characterize a post-processing only optimization. 
Indeed, R-PGPR often reported the same linking interaction across recommended products (low LID). 
This effect could be exacerbated by close temporal proximity the user interactions in ML1M.
Conversely, the combination of in- and post-processing optimization approaches (R-(IP)PGPR) showed that it is possible to mitigate this effect (increase LID) while even improving LIR.

\textbf{Popularity.} 
Figure~\ref{fig:case-study-popularity} depicts the recommendations and the resulting reasoning paths once SEP was optimized. 
In all the considered settings, our approaches resulted in a higher SEP and, therefore, textual explanations mentioning more popular shared entities. 
Differently from the recency perspective, the differences in SEP were higher between settings, with the combined setting showing an evident gain with respect to the other ones.
It should be also noted that, in this case, the diversity of the shared entity (SED) was generally high.
Though only user shared entities were often selected, none of them was repeated multiple times in the same recommended list. 
With this example, we could also appreciate the positive interdependence between SEP and PTD. 
Indeed, when the in-processing was part of the optimization setting (P-(I)PGPR and P-(IP)PGPR), one and two entity types different than the user type appeared in the selected reasoning paths.  

\vspace{1mm}\noindent\textbf{Diversity.} 
Figure~\ref{fig:case-study-diversity} shows the recommended products and selected paths obtained after optimizing for PTD. 
In the provided example, the diversity of explanation path type was very low ($0.10$). 
The most frequent path type was "watched", confirming our initial motivation on the majority of explanations being connected to what other users have experienced (collaborative-filtering like).
In- and post-processing optimization approaches alone were able to improve PTD, but the absolute value was still relatively low. 
In particular, it should be noted that at most two path types different than "watched" were included and linked to the lowest ranked products. 
Such behavior might be due to the way the optimization approaches were formulated. Specifically, once the sets of products to recommend and their corresponding reasoning paths were selected (after the optimization), the products were ranked according to their relevance for the user. 
To let other reasoning path types appear in the top-$n$ recommendations, it was often necessary to introduce products not originally ranked at the top by the recommendation models. Therefore, once finally ranking based on relevance, the newly-added products linked to different path types ended up being at the bottom.
Combining both approaches made it possible to increase the total number of path types different than "watched" to four and partially counter the abovementioned issue, with one of the four linked to the product at position $2$.
However, the majority of path types was still "watched" (low PTC). 
Future work should investigate novel ways of dealing with this issue.

\hlbox{Observation 3}{
As exemplified in the provided use case, compared to the other settings, the combination of both in- and post-processing not only generally reported the highest reasoning path quality scores (explanations linked to more recent user interactions, more popular shared entities, and more diverse reasoning path types) but also showed the highest diversity of linking interactions and shared entities as well as the lowest concentration of path types.}
}

\section{Conclusions and Future Work}\label{sec:conclusions}
In this paper, we conceptualized, assessed, and operationalized three metrics to monitor reasoning path quality at user level. We proposed a suite of in- and post-processing approaches to optimize for these metrics. 
Experiments on three real-world data sets from different domains showed that, compared to the baselines, not only the proposed approaches improve the overall path quality, but also preserve recommendation utility.

Our findings in this study, paired with its limitations, will lead us to extend and operationalize a larger space of user-level reasoning path properties deemed as important for the next generation of explainable RSs. 
Indeed, the proposed properties are not exhaustive by any means, and further studies will be conducted, also via additional user surveys. 
Traditional beyond-accuracy metrics explored in RS research, e.g., serendipity, diversity, and novelty, can be further elaborated in the context of reasoning paths. 
Other properties can be also used to control the fair exposition of the entities pertaining to humans (e.g., producers and actors in the movie domain) in the explanations. 
Another interesting direction can investigate model agnostic explanation subsystems able to turn explanation scores returned by regularized-based explainable RSs into explanation paths to provide textual explanations. 
This would also serve to assess the transferability of our approaches to a larger set of models. 
On the other hand, assessing generalizability to other domains (e.g., education) will require to extend existing data sets with their KG.  
Finally, in the long term, the impact of explainable RSs on the platform and its stakeholders will be evaluated via online experiments at scale.

\appendices
\counterwithin{figure}{section}
\counterwithin{table}{section}

\section{}
To show that the identified reasoning path quality metrics might have an impact on the resulting explanations, we explored the distribution of the interactions per user over time, the distribution of the popularity over the entities, and the distribution of relations per type over the KG, on the three considered data sets (ML1M, LASTFM, and CELL).  

\begin{figure}[!b]
\vspace{6mm}
\centering
\includegraphics[width=.99\linewidth]{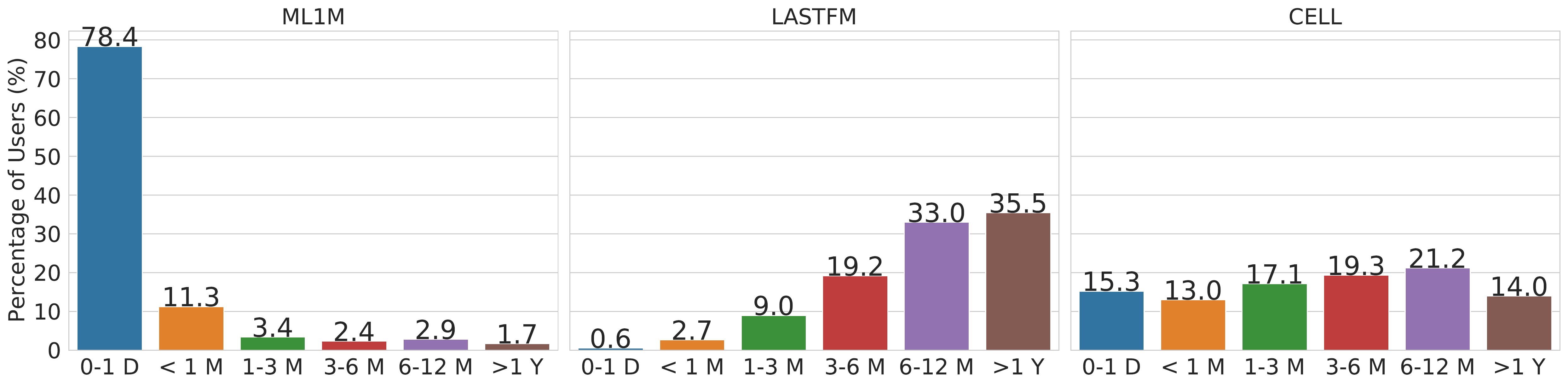}
\caption{For each data set, the percentage of users grouped by distance between user most recent interaction and their other interactions (D = Days, M = Months, Y = Year).} 
\label{fig:average-MRI-distance}
\end{figure}

\vspace{1mm}\noindent\textbf{Recency.} First, we considered that the LIR is defined according to the time the user interacted with the already experienced product. 
It was therefore important to explore how the interactions per user are distributed over time.
To this end, for each user, we computed the average distance in time between their most recent interaction and any other interaction that user performed.
Figure \ref{fig:average-MRI-distance} collects the percentage of users according to their average distance in time over a daily, monthly, or yearly period. 
Most of the users (78.4\%) in ML1M performed their interactions within one day. 
This behavior might be attributed to the nature of the data, collected from a platform where users were explicitly asked to provide ratings for movies.  
Optimizing for LIR would not affect much the temporal perception the user has towards the linking interaction. 
Conversely, users had a tendency to listen to songs over longer time frames (more in the past) in LASTFM.
In Amazon CELL, there was a more balanced distribution of user interactions over time frames. 
For the latter two data sets, optimizing for LIR would produce explanations that are more relevant to the recent interactions and tastes of users, especially for those who tended to interact less frequently over time (right bars).  

\begin{figure}[!t]
\centering
\includegraphics[width=0.5\textwidth]{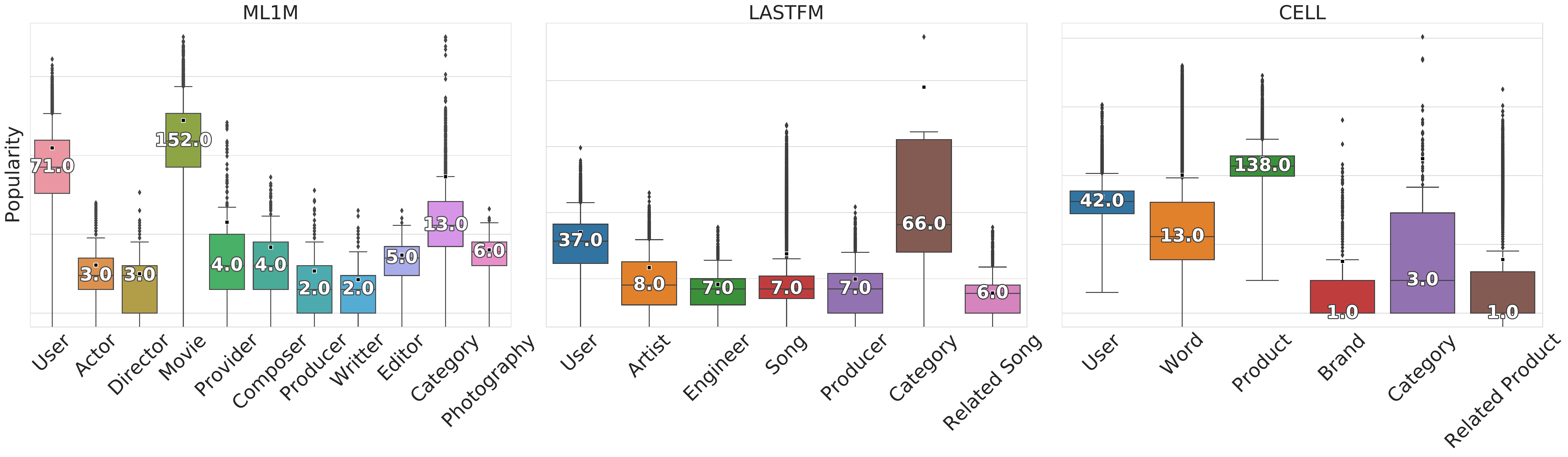}
\caption{For each data set, entity popularity distribution per type (log scale).} 
\label{fig:pop-distr}
\vspace{4mm}
\end{figure}

\begin{figure}[!t]
\centering
\includegraphics[width=0.5\textwidth]{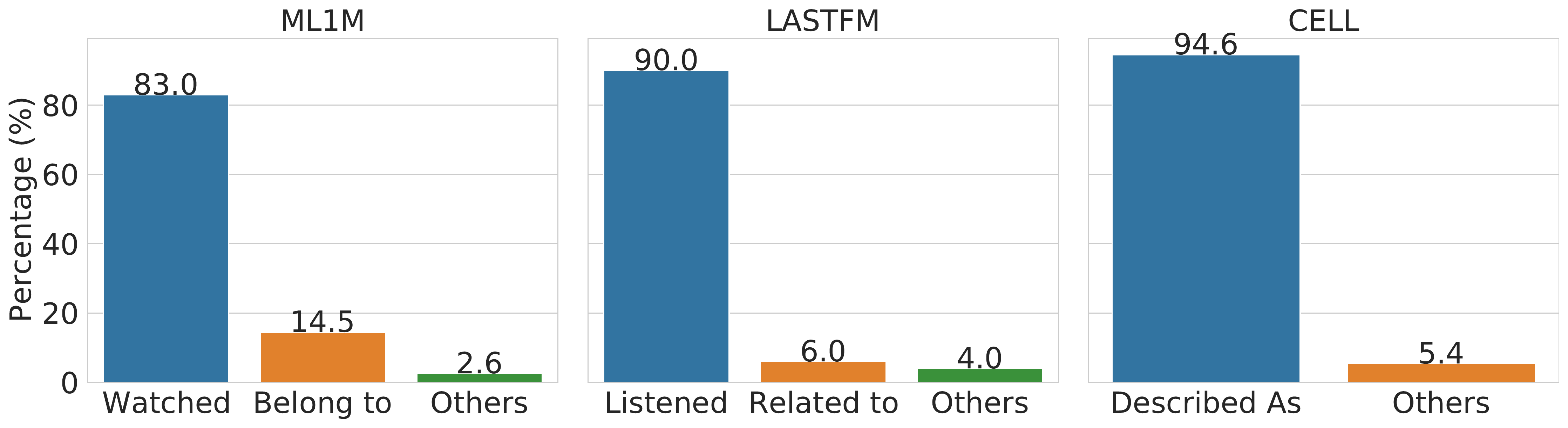}
\caption{Path type (relations type) distribution in the KG.} 
\label{fig:path-unbalance}
\end{figure}

\vspace{1mm}\noindent\textbf{Popularity.} Second, we considered that the SEP is defined according to the popularity of the shared entity included into the reasoning path. 
Figure \ref{fig:pop-distr} depicts how the entity popularity is distributed over entities of a given type.
In ML1M, as expected, users and movies appear as the entities with the highest popularity estimates in the KG. 
Overall, the popularity variation over entities of a given type seemed large and comparable across types. 
In the other two data sets, certain entity types showed a larger variation in popularity than others.
The variation in popularity was substantial in most of the entity types. 
Optimizing for SEP would hence produce explanations that include more popular shared entities.
Given the results collected in Figure \ref{fig:pop-distr}, playing with this property would possibly have a more prominent impact on the shared entities in the selected reasoning paths. 

\vspace{1mm}\noindent\textbf{Diversity.} Finally, we considered that the PTD is defined according to the reasoning path types covered in the recommended list. 
Figure \ref{fig:path-unbalance} shows the distribution of relation types in the KG.
It should be noted that the reasoning path type diversity is limited to the relations available in the KG.
In all data sets, most of the relations came from user interactions (i.e., watched, purchased, described as).
Only a minority of them were derived from the external knowledge in the KG.
It follows that this imbalanced setting may directly influence the path selection step during training and, consequently, the PTD in the recommendations.

\balance
\bibliographystyle{acm} 
\bibliography{cas-refs}

\end{document}